\documentclass[10pt,twocolumn]{article}
\usepackage{authblk}
\usepackage{url}
\usepackage{wrapfig}
\usepackage{color}
\usepackage{graphicx}
\usepackage{amsmath,amsfonts,amssymb,amsthm}
\usepackage{wrapfig,comment}
\usepackage{mathdefs}
\usepackage{proposal}
\usepackage{multirow}
\usepackage{my_macros}
\usepackage{float}
\graphicspath{{figures/}}
\usepackage[justification=justified,font=small]{caption}
\usepackage{subcaption}
\usepackage{graphicx}
\usepackage{units}
\usepackage{upgreek}
\usepackage{lipsum}
\usepackage[export]{adjustbox}
\usepackage{placeins}
\usepackage{cite}
\usepackage[a4paper, total={7in, 9in}]{geometry}
\usepackage{subfiles} 

\title{ Fluorescent  wavefront shaping using incoherent iterative phase conjugation}

\author[1]{Dror Aizik}
\author[2]{Ioannis Gkioulekas}
\author[1]{Anat Levin}
\affil[1]{ Department of Electrical and Computer Engineering, Technion, Haifa, Israel}%
\affil[2]{Robotics Institute, Carnegie Mellon University, PA, USA}
\date{}   
\begin{document} 
	\maketitle

	\begin{abstract}
		Wavefront shaping correction makes it possible to image fluorescent particles  deep inside scattering tissue. This requires determining a correction mask to be placed in both excitation and emission paths. Standard approaches select correction masks by optimizing various image metrics, a process that requires capturing a prohibitively large number of images. To reduce acquisition cost, iterative phase conjugation techniques use the observation that the desired correction mask is an eigenvector of the tissue transmission operator. They then determine this eigenvector via optical implementations of the power iteration method, which require capturing orders of magnitude fewer images.
		Existing iterative phase conjugation techniques assume a linear model for the transmission of light through tissue, and thus only apply to fully-coherent imaging systems. We extend such techniques to the incoherent case for the first time. The fact that light emitted from different sources sums incoherently violates the linear model and makes linear transmission operators inapplicable. We show that, surprisingly, the non-linearity due to incoherent summation results in an order-of-magnitude acceleration in the convergence of the phase conjugation iteration.

	\end{abstract}

	\vspace{-0.1in}
	\section{Introduction}
	
	One of the core challenges when performing linear fluorescence microscopy inside tissue is the fact that biological tissue is highly scattering at visible wavelengths. This limits the clinical applicability of linear fluorescence microscopy techniques to thin superficial layers, as  incoming and outgoing light propagating through the tissue is highly aberrated. In turn, this precludes widespread clinical use for tasks such as vasculature imaging, laser light therapy, and tumor detection.
	
	A promising approach for overcoming the multiple scattering challenge is wavefront shaping correction: if one reshapes the incoming (or outgoing) coherent wavefront, such that its aberration is conjugate to the aberration that will happen inside the tissue, then after propagation the wavefront will focus into a sharp spot inside the tissue. 
	
	Adaptive optics techniques~\cite{Booth2014,Ji2017review,HampsonBooth21review} were first used to correct modest aberrations, for example due to imperfect optics or refractive index variations in the tissue. 
	More recently, wavefront shaping techniques~\cite{YU2015632,Gigan22} have shown that it is possible to focus light through thick, highly-scattering layers~\cite{Vellekoop:07,Yaqoob2008,Vellekoop2010,Vellekoop2012}.

	Wavefront shaping ideas have  found applications in a wide range of imaging modalities, including sound and light, coherent imaging and OCT, and incoherent fluorescence imaging using single-photon and multi-photon excitation. Our  interest in this work is wavefront shaping for linear, single-photon fluorescence feedback.

	%
	
	The practical application of wavefront shaping is hindered by the difficulty of finding the wavefront correction to apply. This wavefront correction varies between different tissue layers, and even between different positions inside the same tissue sample.
	The simplest approach for finding the wavefront correction is to use a so-called guide star~\cite{Horstmeyer15,Tang2012,Katz:14,Wang20142PAdaptive,Liu2018,Fiolka:12,Jang:13,Xu11,Wang2012,Kong:11}: In this case, scattering arises from a strong \emph{single} point source inside tissue, and a wavefront sensor~\cite{Vellekoop2012,Liu2018} directly measures the scattered wavefront. 

	Finding a wavefront shaping correction in the presence of multiple sources\Marina{you mean non coherent?}\Anat{no. this is a general statment and applies both to coherent and incoherent cases} is more challenging, and typically involves optimization strategies relying on a variety of feedback mechanisms~\cite{Li:15,Tang2012,Katz:14,Wang20142PAdaptive,Boniface:19,Boniface2020,Bonora:13,Antonello:20,YeminyKatz2021,Stern:19,Daniel:19,Verstraete:15,Choi2015,Vellekoop:07,Conkey:12,PopoffPhysRevLett2010,Vellekoop2010,Yaqoob2008,chen20203PointTM}.
	This optimization is tractable when the wavefront correction can be described by a small number of parameters (e.g., using Zernike polynomials~\cite{Booth2002,Park2017}).  
	However, to focus inside thick highly-scattering media, it is desired to use all the degrees of freedom of a modern spatial light modulator (SLM), often in the megapixel range. This is posing non-trivial optimization challenges~\cite{Conkey:12,Katz:14,Vellekoop:07,Popoff2011}. Even if we can test every such free parameter only once~\cite{chen20203PointTM}, the very large number of images captured for optimization limits any real-time applicability.

	For fully coherent imaging systems, an alternative class of techniques estimating the wavefront correction is \emph{iterative phase conjugation}. These techniques use the observation that a wavefront shaping  correction focusing on a single point inside tissue is an \emph{eigenvector} of the transmission matrix of the scattering sample~\cite{Popof2011}. They then find these eigenvectors using an optical implementation of the power method~\cite{trefethen97}, which iterates between sending in a wavefront, measuring the scattered wavefront, and using the measurement as the successive input. Often this  procedure converges after a very small number of iterations, leading to an order-of-magnitude acquisition speedup compared to standard optimization approaches. 
	Iterative phase conjugation has found successful applications for sound~\cite{doi:10.1121/1.412285,doi:10.1121/1.424648} and acousto-optics~\cite{Meng2012,Yang2014}, where the propagation is fully coherent.
	Although not presented this way, a similar iterative scheme was also applied for  two-photon fluorescent imaging~\cite{Papadopoulos16}.

	%
	%

	An important assumption underlying the coherent iterative phase conjugation scheme is that light scatters only once.
	This greatly limits its applicability to thin or sparse volumes.
	Our goal in this work is  to develop an iterative phase conjugation approach that is applicable to linear (single-photon) fluorescent imaging.
	As  the emitted light does not excite the tissue or the particles again, by working with fluorescent sources  we can greatly relax the single scattering assumption, making our approach applicable  to much thicker volumes, in particular tissue. 
	
	The primary technical challenge in this setting is that any uncorrected incident wavefront (such as the wavefronts used during the power method) will excite more than one fluorescing point inside the tissue sample, and the excited points will emit light that sums \emph{incoherently}. 
	Consequently, we cannot model the relation between input excitation and output fluorescent emission using a linear transmission operator, as fully-coherent iterative phase conjugation techniques do. To overcome this challenge, we analyze the incoherent case, and report two findings: First, we show that the same power method procedure as in the fully-coherent case can be used to recover the correction pattern also in the incoherent case. Second, we show that, whereas for the fully-coherent case the power method converges at an exponential rate, for the incoherent case it converges at a \emph{doubly-exponential} rate. We demonstrate these findings experimentally,  focusing light on fluorescent beads attached at the back of chicken breast tissue layers. Our technique achieves wavefront correction after capturing as few as $10-30$ images, compared to thousands of images captured by existing optimization-based wavefront shaping strategies for fluorescent imaging~\cite{Boniface:19}.
	

	%
	

	\section{Principle}\label{sec:principle}
\begin{figure*}[t!]
	\begin{center}
		\includegraphics[width= 1\textwidth]{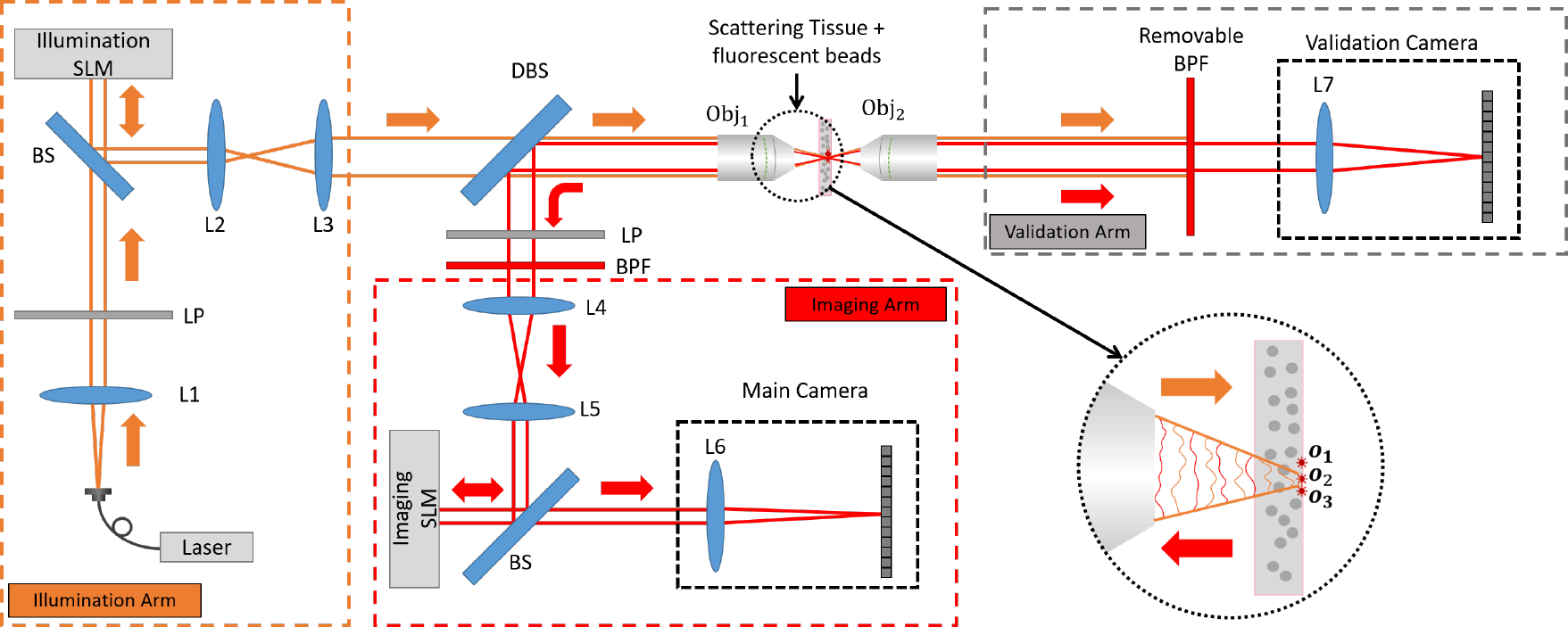}
	\end{center}
	\caption{Our wavefront correction fluorescent microscope setup: A laser beam is exciting fluorescent beads at the back of a tissue layer, and fluorescent emission is scattered again through the tissue, reflects at a dichroic beam-splitter and is collected by a main (front) camera. We place two SLMs in the Fourier planes of both illumination and imaging arms to allow reshaping these wavefronts. A validation camera views the beads at the back of the tissue directly. This camera is  not actually used by the algorithm, and is only  assessing  its success. LP=linear polarizer, BS=beam-splitter, DBS=dichroic beam-splitter, BPF=bandpass filter,  $L1\ldots L7$=lenses, Obj=Objective. }\label{fig:setup}
\end{figure*}

	\figref{fig:setup} shows our imaging setup. A laser beam illuminates a tissue sample via  a microscope objective. A phase SLM at the Fourier plane of the illumination arm modulates the illumination pattern. The modulated laser light excites fluorescent beads at the back of the sample. The emitted light is collected via the same objective, and reflected at a dichroic beam-splitter. A second phase SLM at the Fourier plane of the imaging arm modulates the emitted light. Lastly, the modulated light is measured by the front camera, which captures the images used by our algorithm. 
	The setup includes a second validation camera behind the tissue sample. \blue{  In our experiments we attached fluorescent beads at the back of the tissue layer, so that the validation camera can   image them directly.} We emphasize that measurements from this camera are \emph{not} used by our algorithm, and that we only use the camera for validation purposes, to assess focusing quality and to image an undistorted reference of the bead layout.
	
	We  derive a strategy for efficiently finding a wavefront shaping modulation pattern for the illumination arm, allowing us to focus all light into a single spot inside the tissue sample. Once we have found the modulation pattern, we use the same modulation to also correct the emitted light in the imaging arm. This is possible because, in our linear fluorescent imaging setting, emission and excitation wavelengths are relatively close. Our approach extends to the incoherent imaging case iterative phase conjugation ideas that were previously used with coherent illumination. 
	We begin our presentation by reviewing the coherent case, and then introduce the incoherent one.
	
	\paragraph{Coherent iterative phase conjugation.}
	\blue{Consider a set of $K$ scattering (non fluorescent) particles inside a  sample, and denote their positions by $\ptd_1,\ldots,\ptd_K$.
		We denote by  $\bu$ the value of an incoming 2D electric field at the input plane, and by $\bou$ a $K\times 1$ vector of  the field propagating through the sample at each of the $K$  scatterers. 
		Although $\bu$ is a 2D field, we reshape it as a 1D vector  and relate $\bou$ to $\bu$ as $\bou=\TMi \bu$,} where $\TMi$ is the incoming transmission matrix describing coherent light propagation. $\TMi$ is specific to the tissue sample being tested.  
	Likewise, we denote by $\TMo$ the back-propagation transmission matrix, describing the light returning from the particles to the sensor. 
	%
	The propagation of light to the particles and back to the sensor 
	is then modeled using the \emph{combined transmission matrix} \BE \label{eq:Ta-def} \TMa\equiv\TMo\cdot \TMi.\EE 
	Note that \equref{eq:Ta-def} offers a simplistic description of light propagation, assuming there is not much light back-scattered  from other structures in the medium apart of the listed particles $\ptd_1,\ldots,\ptd_K$, and  multiple scattering  between the particles is negligible. 
	
	Under fully coherent illumination,  the input illumination and the measured speckle intensity are related as
	\BE\label{eq:int-Ta}
	I=|\TMa\bu|^2.
	\EE
	Our goal is to find an illumination pattern $\bu$ that will focus on one of the particles, so that  $\bou$ is a \emph{one-hot vector}---non-zero only at a single point $\ptd_k$ for some $k$ value. We note that focusing at any of the particles is sufficient for our setting; below, we show that once we focus at one point, we can use the memory effect to focus at nearby ones.
	
	To find a wavefront modulation we need access to $\TMi$, but in practice we can only measure $\TMa$. 
	The wave conjugation principle states that the returning transmission matrix is the transpose of the incoming one, $\TMo={\TMi}^\top$~\cite{RevModPhys.89.015005}. With this assumption,  consider an illumination field $\bu$ that, after propagating through the tissue sample, generates  a one-hot  $\bou$ vector. If we focus all light at one particle,  then by the wave conjugation principle the returning field is proportional to the incoming one.
	Therefore, we can express the returning intensity in \equref{eq:int-Ta} as
	\BE
	I=s|\bu|^2=|\TMa\bu|^2,
	\EE 
	where $s$ is a scale factor.
	That is, a focusing wavefront\Marina{maye you mark the focused one with $u_{focus}$} is an {\em eigenvector} of the combined transmission matrix $\TMa$. Consequently, if we can compute eigenvectors efficiently, we can find a wavefront that focuses all the light in a single spot.
	
	A common class of numerical algorithms for computing matrix eigenvectors follows the \emph{power method}~\cite{trefethen97}. This algorithm relies on the fact that the sequence $\bu,\TMa \bu,(\TMa)^2 \bu, (\TMa)^3 \bu\ldots$ converges exponentially-fast to the largest eigenvector of $\TMa$.
	Iterative phase conjugation algorithms~\cite{doi:10.1121/1.412285,doi:10.1121/1.424648,Meng2012,Yang2014} do not acquire the full transmission matrix $\TMa$, but instead directly measure its optical operation on wavefronts of interest.
	They begin by illuminating the sample with a random wavefront $\bu^0$, then iteratively measure the resulting output wavefront, and use its conjugate as a successive illumination pattern. That is, at the $t$-th iteration, the incident wavefront is $\bu^{\blue{(t)}}=(\TMa \bu^{\blue{(t-1)}})^*$, where $^*$ denotes complex conjugation. 
	When measuring  intensity images $I=|\TMa \bu^{\blue{(t)}}|^2$, computing $\bu^{\blue{(t+1)}}$ also requires estimating the phase of the measured intensity pattern.
	
	Using the exponential convergence property of the power method it can be shown~\cite{trefethen97}, and we review the derivation in the supplement,  that the energy focused on the $k$-th particle at the $t$-th iteration follows a geometric sequence of the form 
	\BE\label{eq:exp-conv-coherent-main}
	|\bou^{\blue{(t)}}_k|=\frac{1}{N_t} \lambda_k^t \cdot c_k,
	\EE
	for constants $\lambda_k,c_k$ and a normalization factor $N_{\blue{(t)}}$ we derive in the supplement.
	\equref{eq:exp-conv-coherent-main} implies that the energy at the $k$-th particle scales exponentially with the iteration number $t$. Thus, each iteration increases the gap in energy between the strongest and second strongest particles, and the sequence quickly converges to a one-hot $\bou$ vector. 
	
	\paragraph{Incoherent phase conjugation.}
	The main limitation of coherent iterative phase conjugation is that to describe the propagation using the model of \equref{eq:Ta-def} one neglects multiple scattering between the particles, as well as back-scattering from any other tissue components. This in turn limits the applicability of the technique to thin or sparse volumes. By using fluorescent emission we remove this restriction, because even in thick tissue it is reasonable to assume that the emitted light does not excite the tissue or the other beads again. Moreover, we show that the incoherent summation of fluorescent emission results in largely accelerated convergence.  
	However, an adaptation of the power method to the incoherent case is not straightforward due to the non-linearity imposed by incoherent emission. 

	To study the incoherent case we need to adjust the above model  in two ways. First, we now mark by  $\ptd_1,\ldots,\ptd_K$ the positions of the fluorescent particles rather than all scatterers in the volume. We use $\TMi$ to describe propagation at the excitation wavelength $\lambda_i$, and $\TMo$ to describe propagation at the emission wavelength $\lambda_o$. Despite the small difference between emission and excitation wavelengths, we still assume that $\TMo\approx {\TMi}^\top$. Note that $\TMi,\TMo$ describe multiple scattering events by other tissue components apart of the listed fluorescent particles $\ptd_1,\ldots,\ptd_K$.

	
	Second, whereas in the coherent case the output wavefront is a linear function of the input, $\TMa \bu^{\blue{(t)}}$, this linear model no longer holds when incoherently summing light from different emitters. To derive an image formation model for this case, we again use $\bou=\TMi \bu$  to denote the field arriving at the fluorescent emitters. 
	Fluorescent emission is proportional to the intensity of $\bou$, and the recorded intensity equals an incoherent summation 
	\BE \label{eq:in-img-form} 
	I=\sum_k |\ckTMo|^2 |\bou_{\blue{k}}|^2,
	\EE
	where $\ckTMo$ is the $k$-th column of $\TMo$.
	
	If we manage to focus and $\bou$ is a one-hot vector, then there is only a single non zero term in the summation of \equref{eq:in-img-form}.
	\blue{Denoting the index of this non zero entry by $k_o$ we can express the intensity in \equref{eq:in-img-form} as
		\begin{equation}
			\begin{gathered}
				I=|\TMo_{:,k_o}|^2 |\bou_{k_o}|^2=|\TMo_{:,k_o}\bou_{k_o}|^2=|\TMo\bou|^2=\\=|\TMo\TMi\bu|^2= |\TMa \bu|^2.	
			\end{gathered}
		\end{equation}
		So effectively, when focusing is achieved, \equref{eq:in-img-form} reduces to \equref{eq:int-Ta},} and the measured intensity is equivalent to $|\TMa\bu|^2$. Therefore in the incoherent case, a focusing wavefront is still an eigenvector of the transmission operator $\TMa=\TMo\cdot\TMi$.
	
	Motivated by this observation we apply iterative phase conjugation as in the coherent case. As we measure only the intensity of the emitted light, to recover the phase of the wavefront, we use a phase diversity acquisition scheme~\cite{MUGNIER20061}. We place $J=5$ known modulation patterns $H^j$ on the phase SLM of the imaging arm. At the $t$-th iteration, we measure speckle intensity images
	\BE
	I^{\blue{(t,j)}}=\sum_k |h^j\star \ckTMo |^2 |\bou^{\blue{(t)}}_{\blue{k}}|^2,
	\EE
	where $\star$ is convolution, $h^j$ is the Fourier transform of the pattern we placed on the SLM, and $\blue{|\bou^t_k|^2}$ is the intensity arriving at the $k$-th particle in the $t$-th iteration.  We use gradient descent optimization to find a complex wavefront $\bu^{(t+1)}$ minimizing
	\BE\label{eq:phase-diversity-opt-main}
	\sum_j \left|I^{\blue{(t,j)}}-|h^j\star \bu^{\blue{(t+1)}}|^2\right|^2.
	\EE
	We then use the conjugate of the estimated wavefront as the excitation of the next iteration, and display it on the SLM of the illumination arm.
	
	When the intensity image is an incoherent summation from multiple sources there is typically no wavefront minimizing \equref{eq:phase-diversity-opt-main} with zero error.
	Despite this, we show in the supplement that the resulting wavefront is approximately equal to a weighted linear combination of the wavefronts $\ckTMo$ generated by the individual sources. Sources with stronger emission receive a higher weight in the reconstruction, which further increases their weight in the next iteration of the algorithm.
	
	
	In the supplement, we analyze the differences between the coherent and incoherent models, and we show that the incoherent summation results in an asymptotically faster convergence rate. In particular, we prove the following claim.
	\begin{claim}\label{clm:doubly-exp-incoherent}
		The convergence of the power iterations in the incoherent case follows a doubly exponential sequence of the form
		\BE
		|\bou_{\blue{k}}^{\blue{(t)}}|^2=\frac{1}{N_{\blue{(t)}}} (\lambda_k)^{2^t} \cdot c_k
		\EE
		for scalars  $\lambda_k,c_k$ derived in the supplement. 
	\end{claim}
	To understand the difference, we note that in the coherent case of \equref{eq:exp-conv-coherent-main}, the energy at the different particles scales as $\lambda_k^t$. In the incoherent case, we get another exponential factor, and energy scales as $ (\lambda_k)^{2^t} $. Intuitively, this is because the fluorescent emission is proportional to the intensity of the field $|\bou^{(t)}|^2$ arriving at the particles, rather than to the field $\bou^{(t)}$ itself. \blue{ As $\bou^{(t)}$ is squared {\em in every iteration}, the squaring is accumulated into another exponential term.  }
	
\begin{figure}[t!]
	\begin{center}
		\begin{tabular}{@{}c@{~~}c@{}}
			
			\includegraphics[width= 0.23\textwidth]{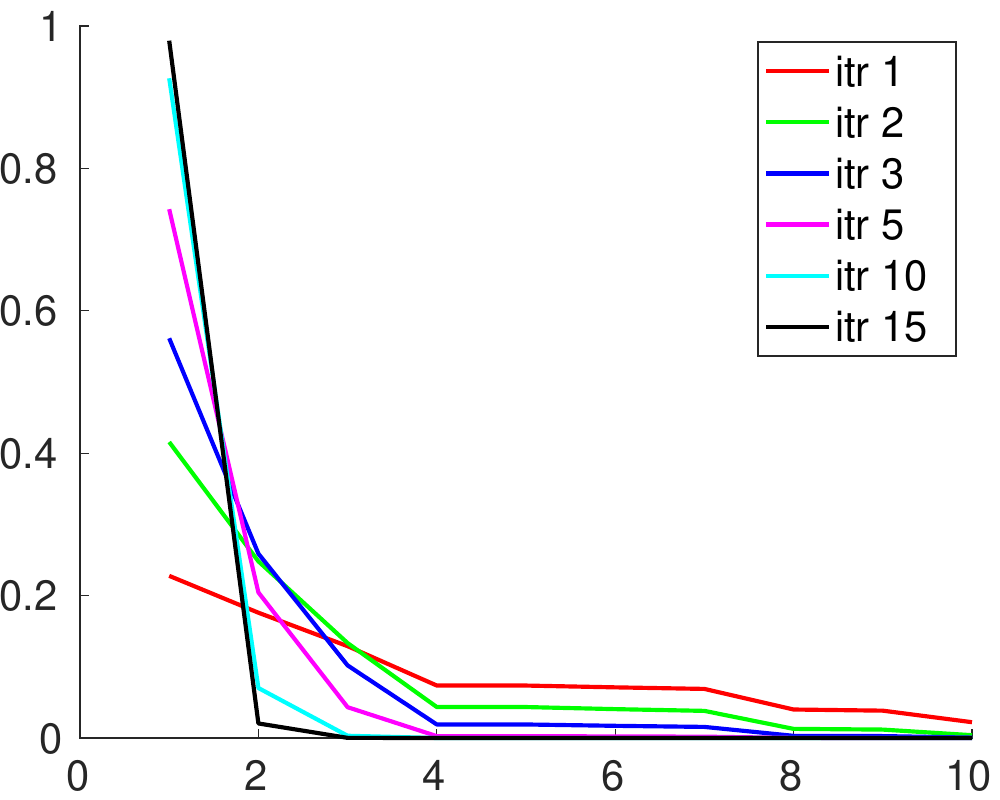}&
			\includegraphics[width= 0.23\textwidth]{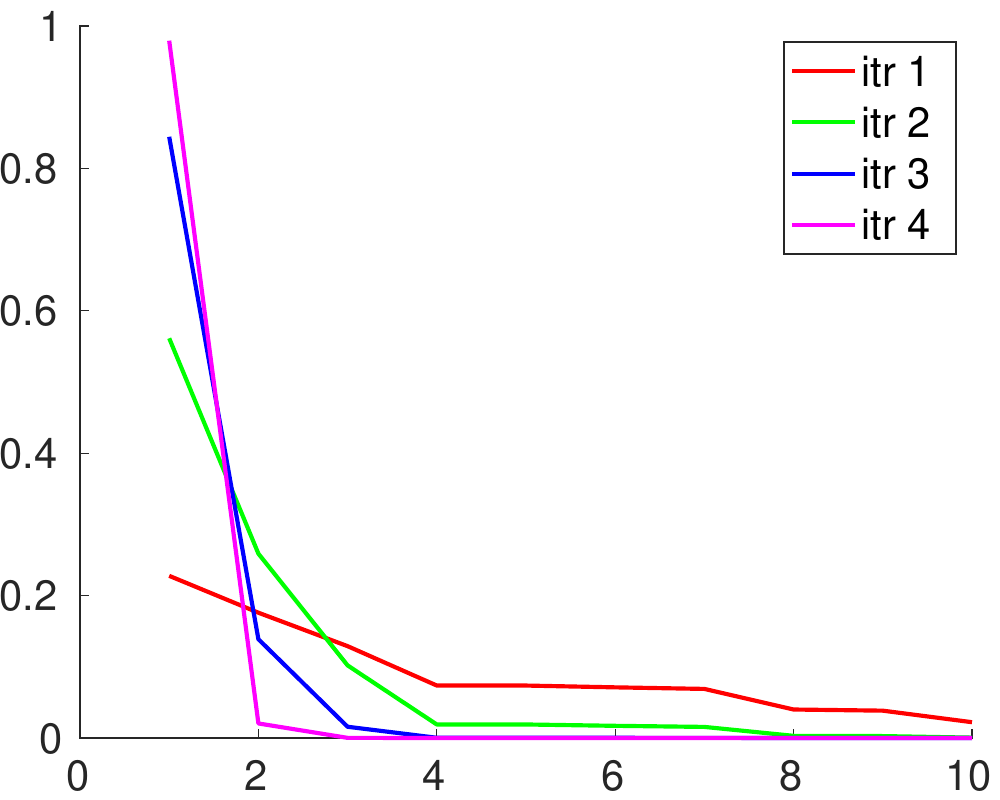}\\
			{(a) Coherent}&{(b) Incoherent} 
		\end{tabular}
		\caption{\blue{Simulating coherent and incoherent convergence: We plot the power of scatterers $|\bou^{(t)}_k|^2$, for different iterations of the iterative phase conjugation algorithm.  As predicted by theory, the incoherent case converges into a one-hot vector within a smaller number of iterations (compare 4 incoherent iterations to 15 coherent ones). The $x$ axis of our plot corresponds to scatterer index $k$, where for ease of visualization we sort these  in decreasing order of power.  	}	}\label{fig:c_ic_convergence}
	\end{center}
\end{figure}
	
	\blue{
		To visualize the faster convergence, in \figref{fig:c_ic_convergence} we simulated  coherent and incoherent power iterations on a random transmission matrix sampled as described in  supplement. }
	
	
	
	In practice, in the hardware implementation described below, our algorithm converged  within about $2-6$ iterations.
	Accounting for the $5$ images used for phase acquisition at each step, our approach can find a wavefront correction pattern using about $10-30$ image measurements. This provides orders of magnitude speedup compared to recent optimization-based approaches recovering a wavefront shaping correction pattern using a single-photon fluorescent feedback, which requires capturing thousands of images~\cite{Boniface:19}.

\begin{figure*}[t!]
	\begin{center}
		\begin{tabular}{@{}c@{~~}c@{~~}c@{~~}c@{~~}c@{~~}c@{~~}c@{~~}c@{}}
			&&Initialization&Iteration 1&Iteration 2&Iteration 3&Iteration 4&Iteration 5\\
			{\raisebox{0.90cm}{\rotatebox[origin=c]{90}{Main Camera }}}&
			{\raisebox{0.90cm}{\rotatebox[origin=c]{90}{No mod.}}}&
			\includegraphics[width= 0.12\textwidth]{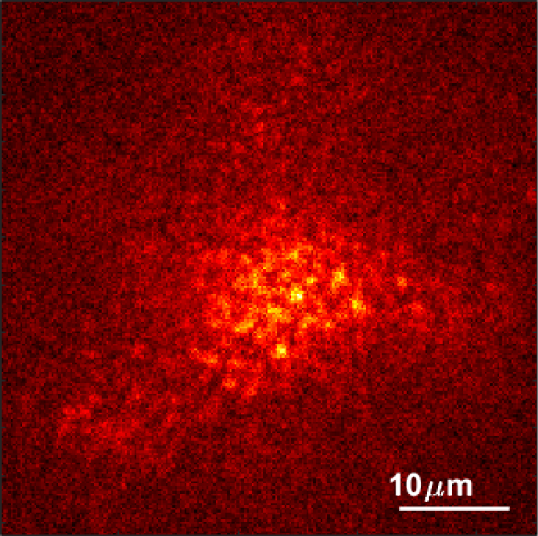}&		
			\includegraphics[width= 0.12\textwidth]{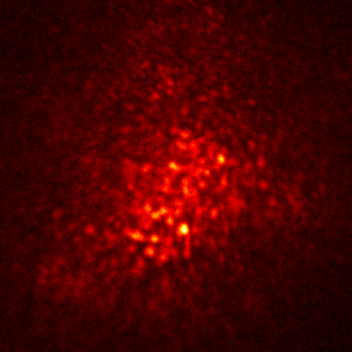}&
			\includegraphics[width= 0.12\textwidth]{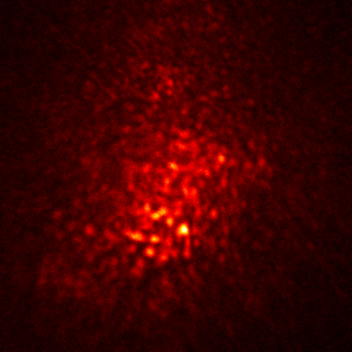}&
			\includegraphics[width= 0.12\textwidth]{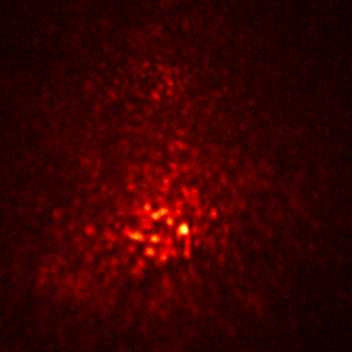}&
			\includegraphics[width= 0.12\textwidth]{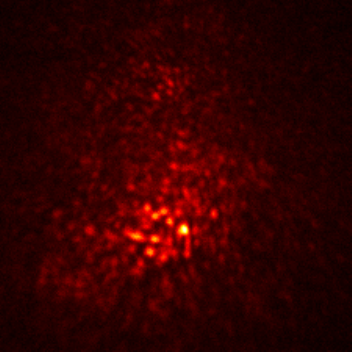}&			
			\includegraphics[width= 0.12\textwidth]{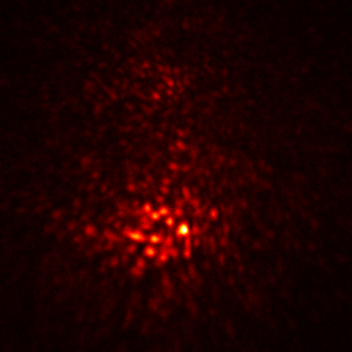}
			\\
			{\raisebox{0.90cm}{\rotatebox[origin=c]{90}{Main Camera }}}&
			{\raisebox{0.90cm}{\rotatebox[origin=c]{90}{with mod.}}}&
			\includegraphics[width= 0.12\textwidth]{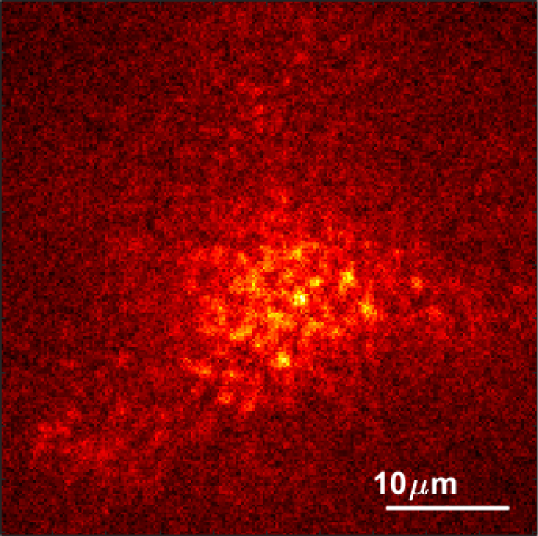}&
			\includegraphics[width= 0.12\textwidth]{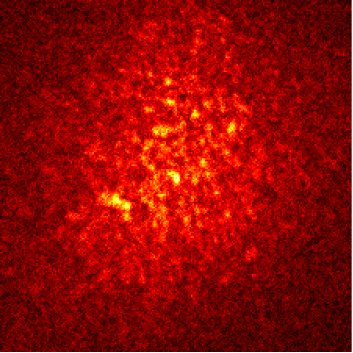}&
			\includegraphics[width= 0.12\textwidth]{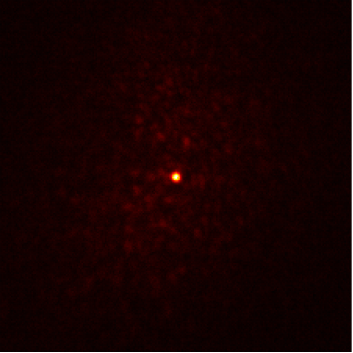}&
			\includegraphics[width= 0.12\textwidth]{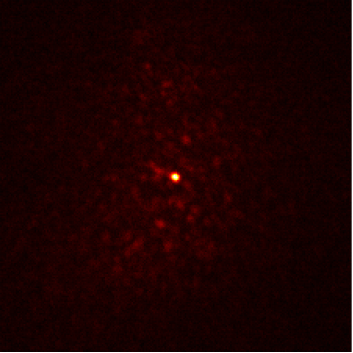}&
			\includegraphics[width= 0.12\textwidth]{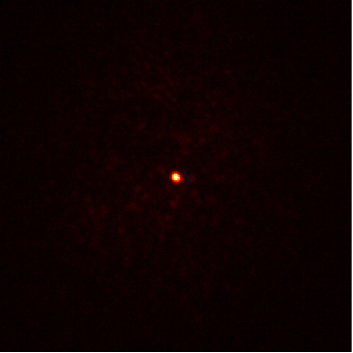}&
			\includegraphics[width= 0.12\textwidth]{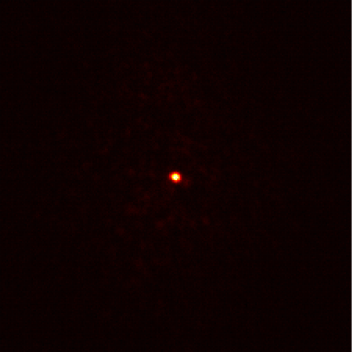}
			\\	
			{\raisebox{0.90cm}{\rotatebox[origin=c]{90}{Val. Camera }}}&
			{\raisebox{0.90cm}{\rotatebox[origin=c]{90}{Emission}}}&
			\includegraphics[width= 0.12\textwidth]{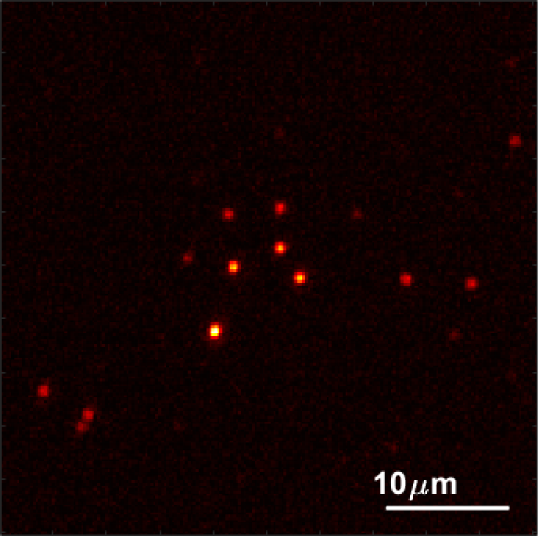}&			
			\includegraphics[width= 0.12\textwidth]{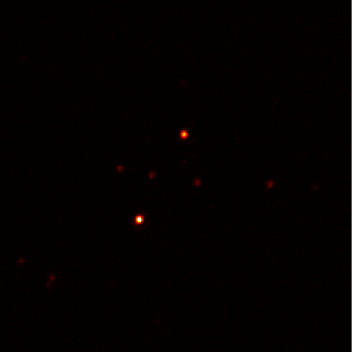}&
			\includegraphics[width= 0.12\textwidth]{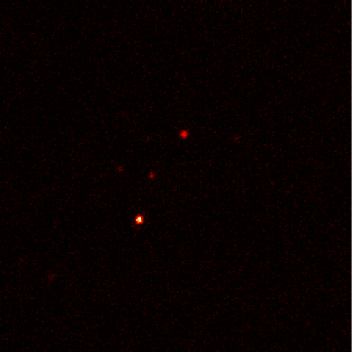}&
			\includegraphics[width= 0.12\textwidth]{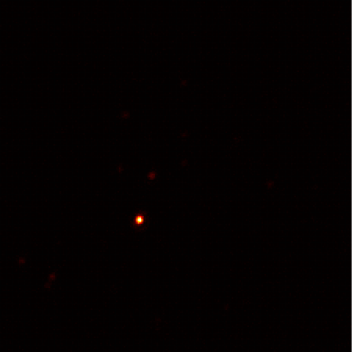}&
			\includegraphics[width=0.12\textwidth]{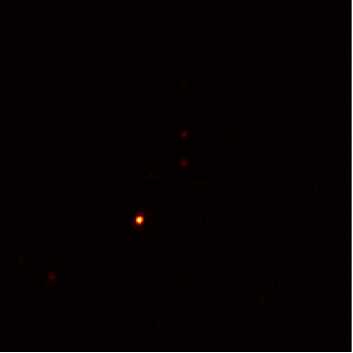}&
			\includegraphics[width= 0.12\textwidth]{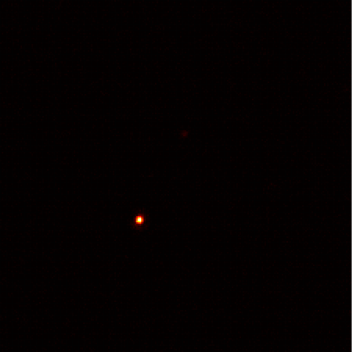}
			
			\\
			{\raisebox{0.90cm}{\rotatebox[origin=c]{90}{Val. Camera }}}&
			{\raisebox{0.90cm}{\rotatebox[origin=c]{90}{Excitation}}}&
			\includegraphics[width= 0.12\textwidth]{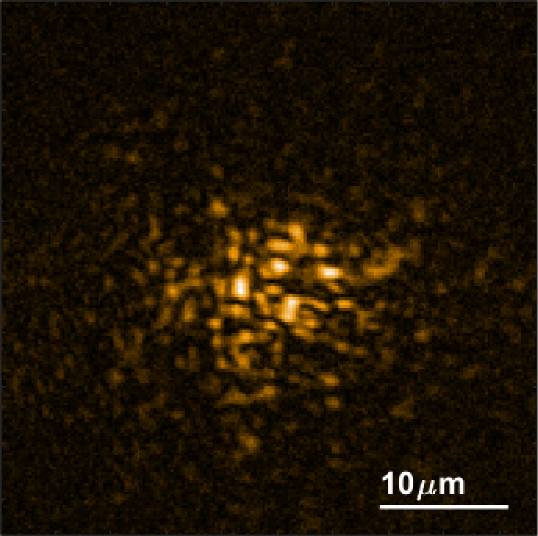}&		
			\includegraphics[width= 0.12\textwidth]{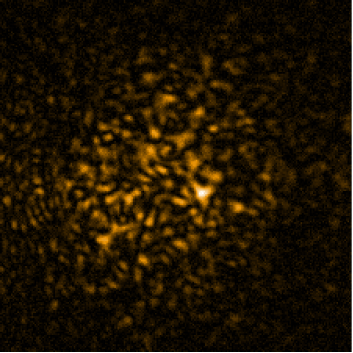}&
			\includegraphics[width= 0.12\textwidth]{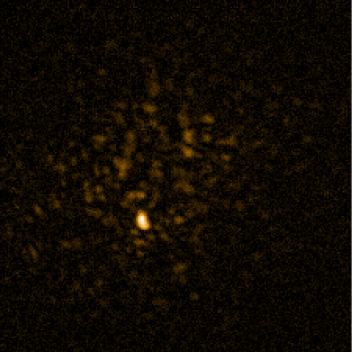}&
			\includegraphics[width= 0.12\textwidth]{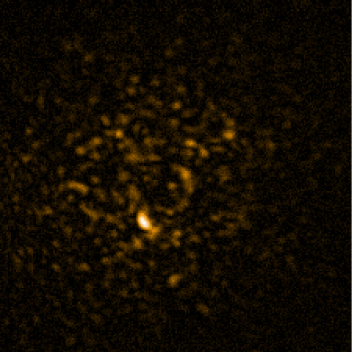}&
			\includegraphics[width= 0.12\textwidth]{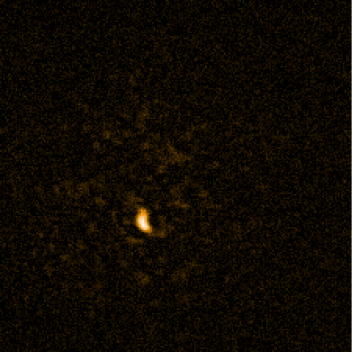}&			
			\includegraphics[width= 0.12\textwidth]{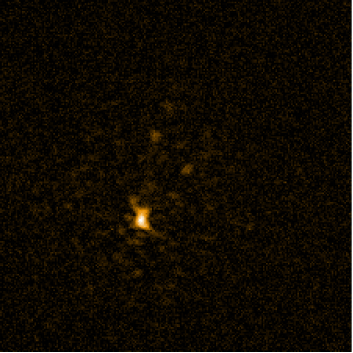}\\

			\\
			{\raisebox{0.90cm}{\rotatebox[origin=c]{90}{Main Camera }}}&
			{\raisebox{0.90cm}{\rotatebox[origin=c]{90}{No mod.}}}&
			\includegraphics[width= 0.12\textwidth]{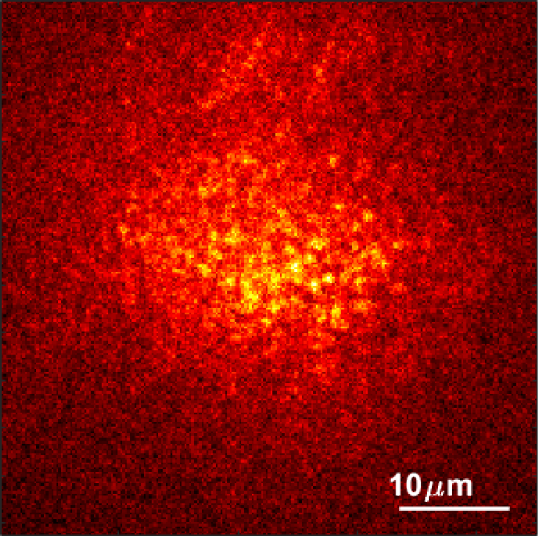}&		
			\includegraphics[width= 0.12\textwidth]{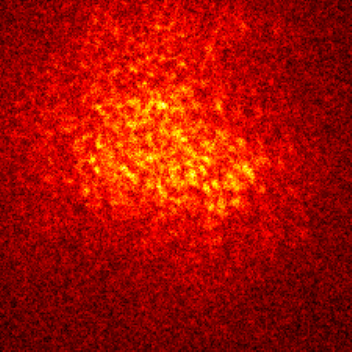}&
			\includegraphics[width= 0.12\textwidth]{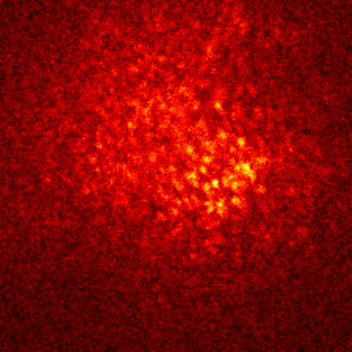}&
			\includegraphics[width= 0.12\textwidth]{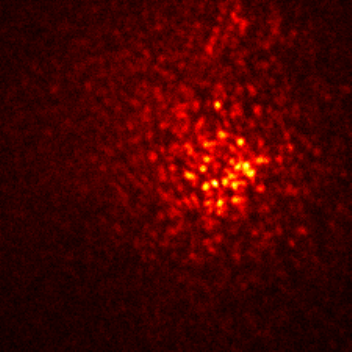}&
			\includegraphics[width= 0.12\textwidth]{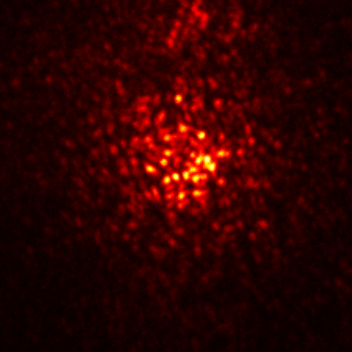}&
			\includegraphics[width= 0.12\textwidth]{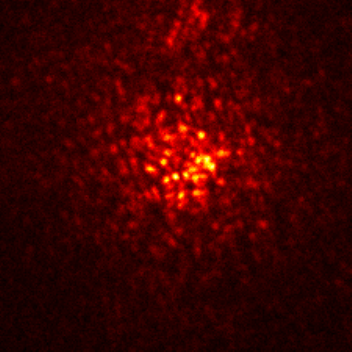}\\
			{\raisebox{0.90cm}{\rotatebox[origin=c]{90}{Main Camera }}}&
			{\raisebox{0.90cm}{\rotatebox[origin=c]{90}{With mod.}}}&
			\includegraphics[width= 0.12\textwidth]{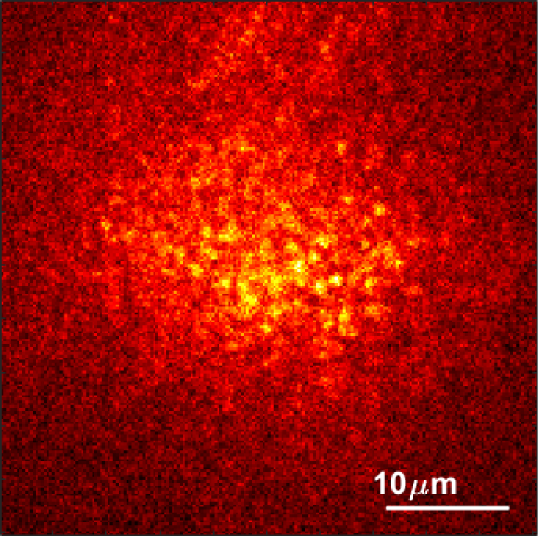}&
			\includegraphics[width= 0.12\textwidth]{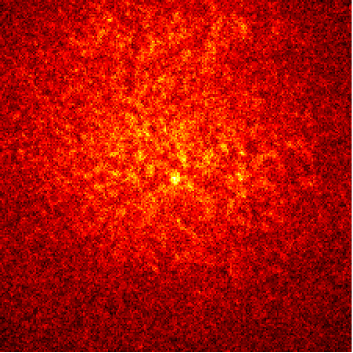}&
			\includegraphics[width= 0.12\textwidth]{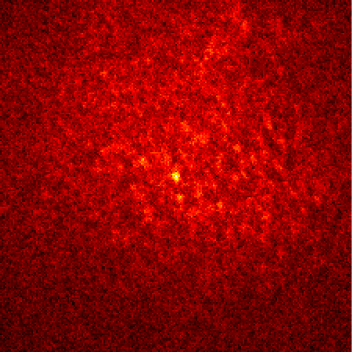}&
			\includegraphics[width= 0.12\textwidth]{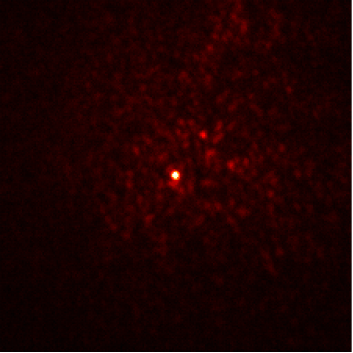}&
			\includegraphics[width= 0.12\textwidth]{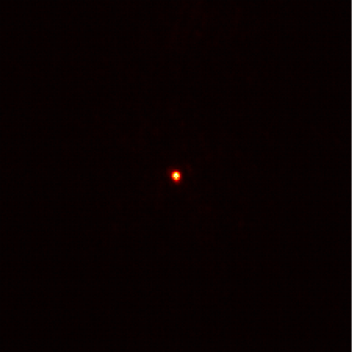}&
			\includegraphics[width= 0.12\textwidth]{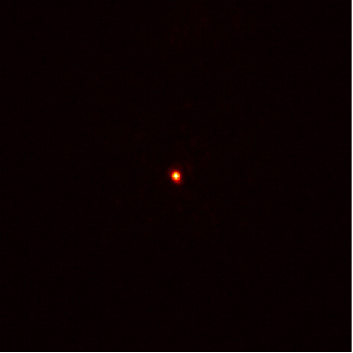}\\
			{\raisebox{0.90cm}{\rotatebox[origin=c]{90}{Val. Camera }}}&
			{\raisebox{0.90cm}{\rotatebox[origin=c]{90}{Emission}}}&
			\includegraphics[width= 0.12\textwidth]{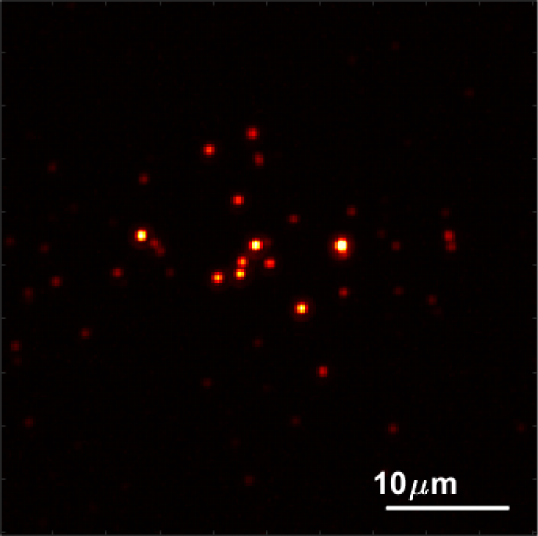}&		
			\includegraphics[width= 0.12\textwidth]{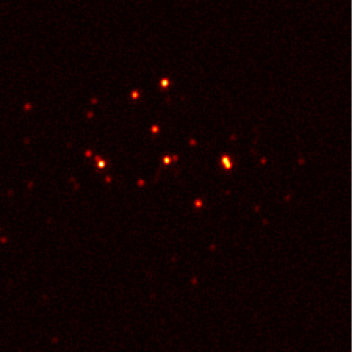}&
			\includegraphics[width= 0.12\textwidth]{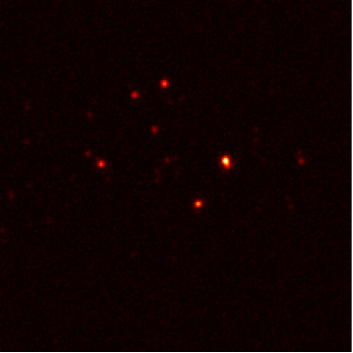}&
			\includegraphics[width= 0.12\textwidth]{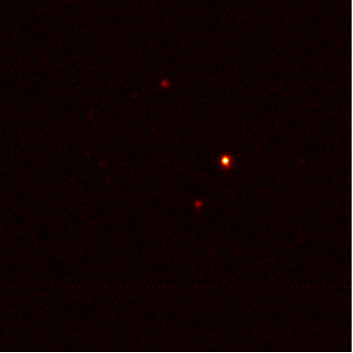}&
			\includegraphics[width= 0.12\textwidth]{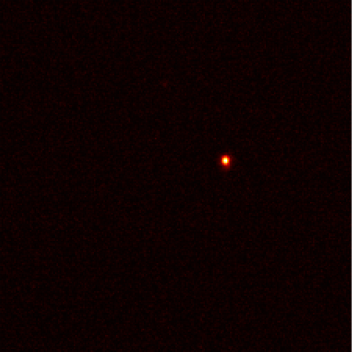}&
			\includegraphics[width= 0.12\textwidth]{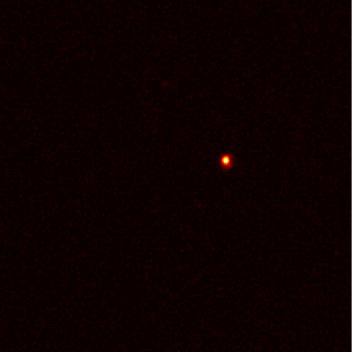}\\
			{\raisebox{0.90cm}{\rotatebox[origin=c]{90}{Val. Camera }}}&
			{\raisebox{0.90cm}{\rotatebox[origin=c]{90}{Excitation}}}&  
			\includegraphics[width= 0.12\textwidth]{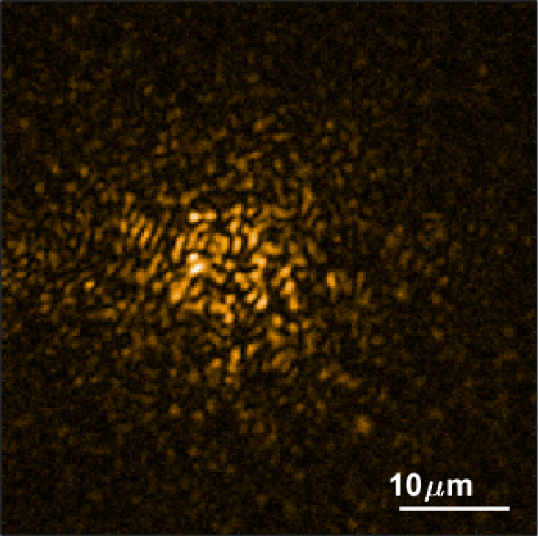}&		
			\includegraphics[width= 0.12\textwidth]{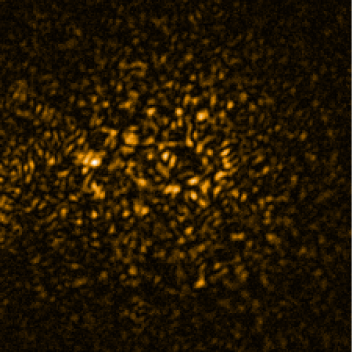}&
			\includegraphics[width= 0.12\textwidth]{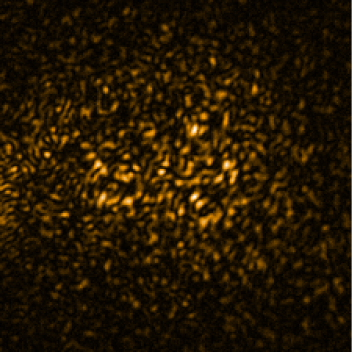}&
			\includegraphics[width= 0.12\textwidth]{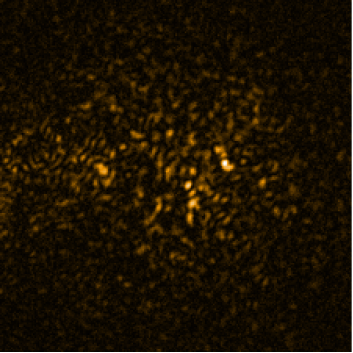}&
			\includegraphics[width= 0.12\textwidth]{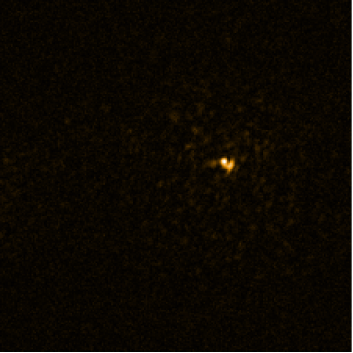}&
			\includegraphics[width= 0.12\textwidth]{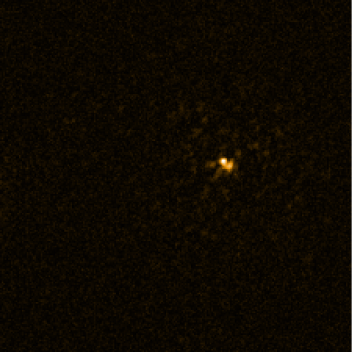}\\

		\end{tabular}
		\caption{	Algorithm convergence. We show the iterations of our power algorithm on two different tissue samples. We demonstrate views via the main camera seeing the front of the tissue with and without the modulation correction, and the validation camera observing fluorescent beads directly. To better appreciate the focusing  we used the validation camera to capture both the excitation and  emission wavelengths. In the first iteration we see a speckle image, but as power iterations proceed the illumination wavefront converges and focuses on a single bead.  When the same modulation pattern is placed at the imaging arm, imaging  aberrations are corrected and one can  see a sharp image of the excited bead. Note that images in different iterations have very different ranges, and for better visualization each image was normalized to its own maximum. }\label{fig:convergence}
	\end{center}
\end{figure*}
\begin{figure*}[h!]
	\begin{center}
		
		\begin{tabular}{@{}c@{~~}c@{~~}c@{~~}c@{~~}c@{}}

			\includegraphics[width= 0.15\textwidth]{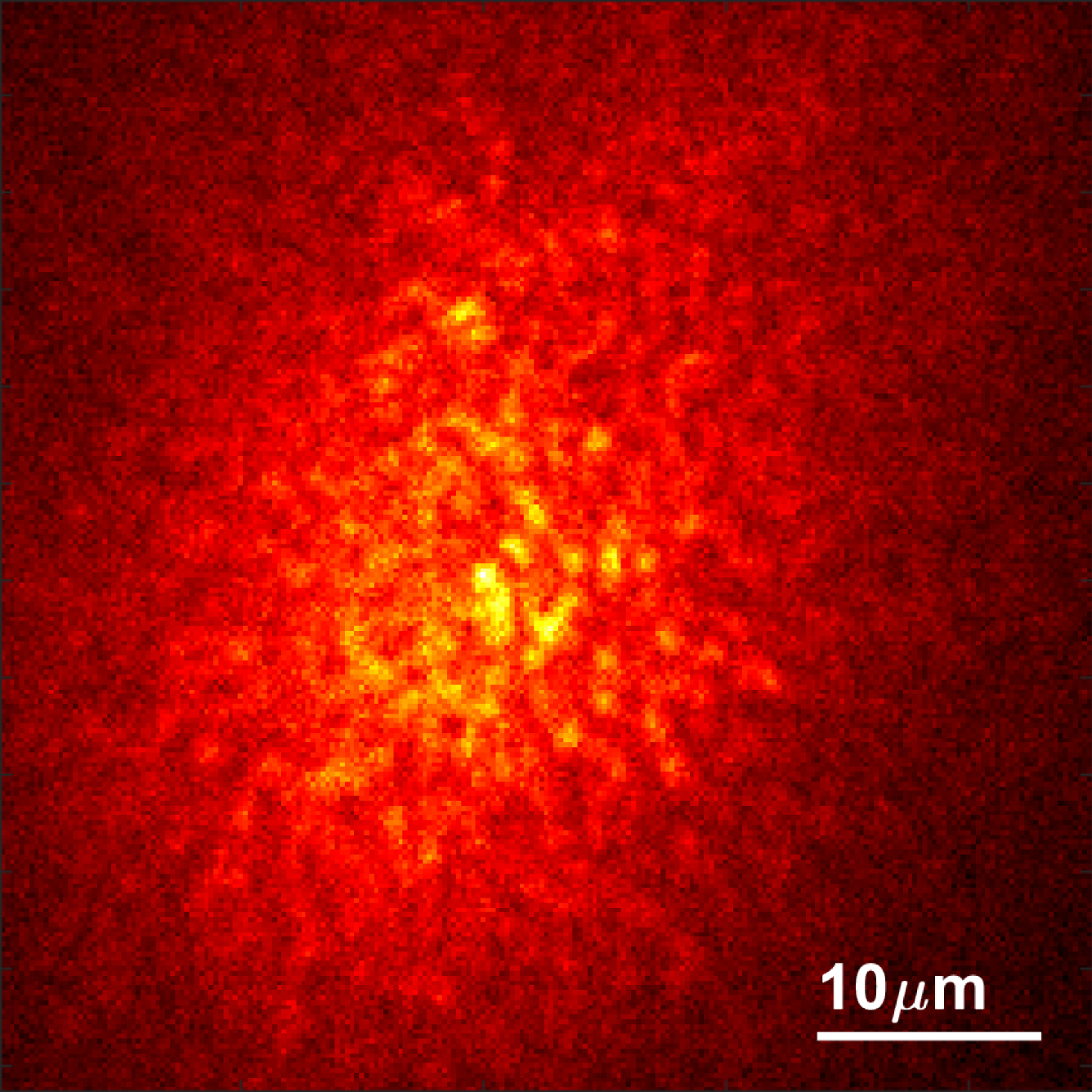}&
			\includegraphics[width= 0.15\textwidth]{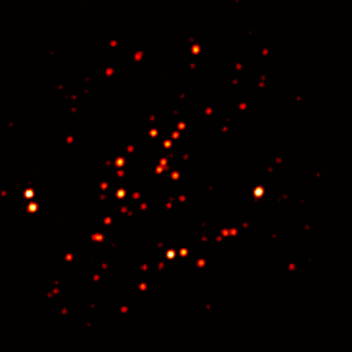}&
			\includegraphics[width= 0.15\textwidth]{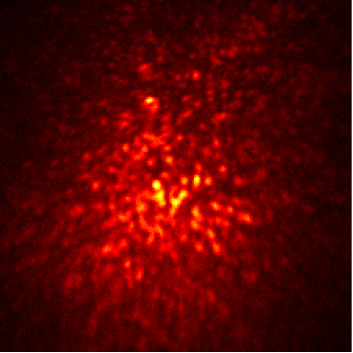}&
			\includegraphics[width= 0.15\textwidth]{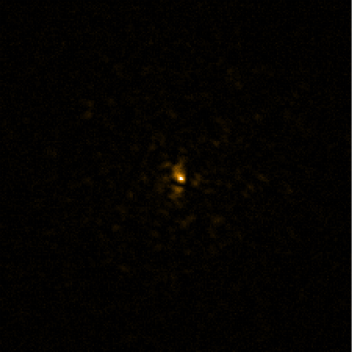}&
			\includegraphics[width= 0.15\textwidth]{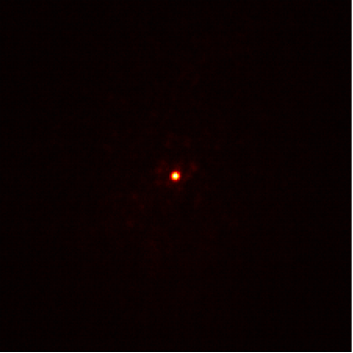}\\

			\includegraphics[width= 0.15\textwidth]{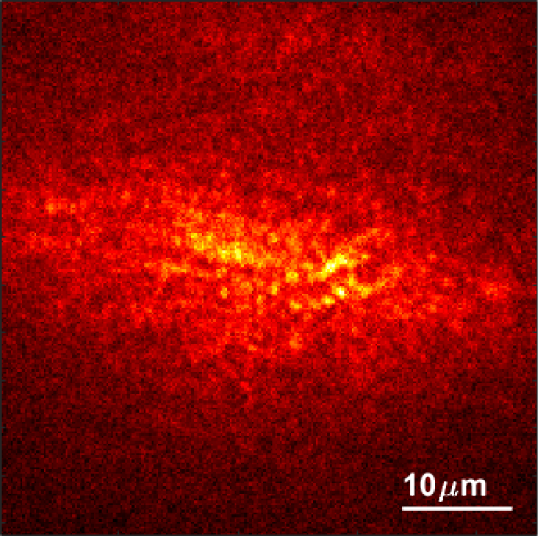}&
			\includegraphics[width= 0.15\textwidth]{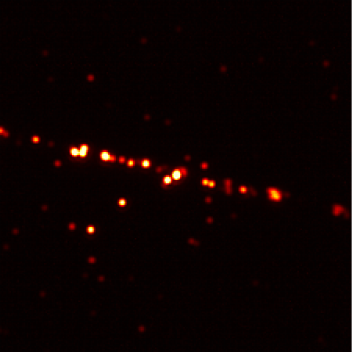}&
			\includegraphics[width= 0.15\textwidth]{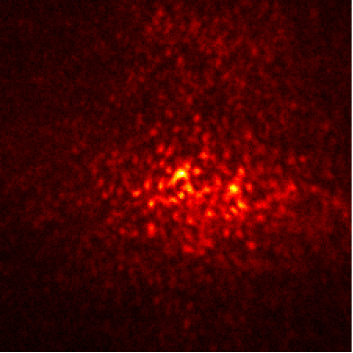}&
			\includegraphics[width= 0.15\textwidth]{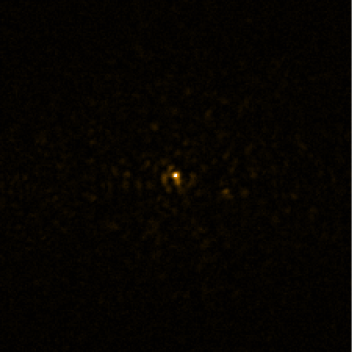}&
			\includegraphics[width= 0.15\textwidth]{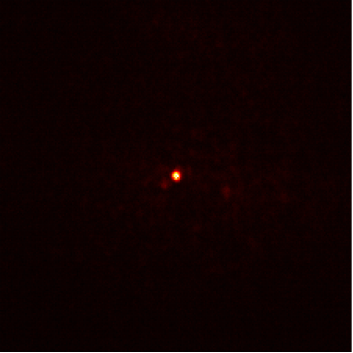}\\

			\includegraphics[width= 0.15\textwidth]{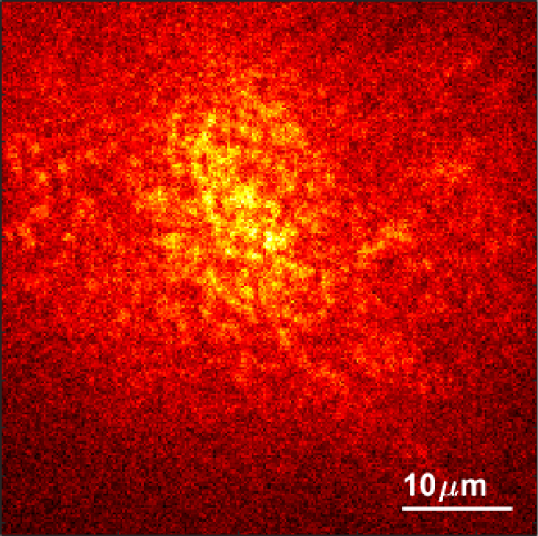}&
			\includegraphics[width= 0.15\textwidth]{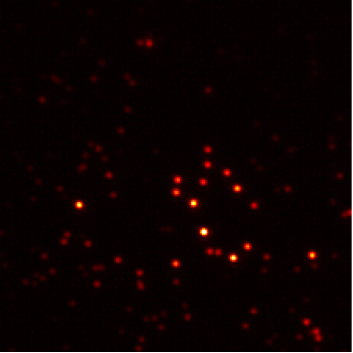}&
			\includegraphics[width= 0.15\textwidth]{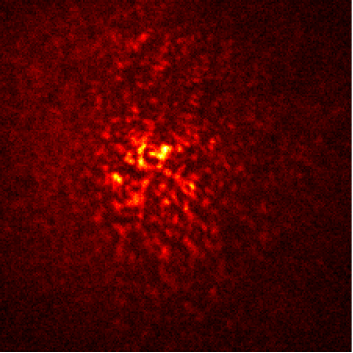}&
			\includegraphics[width= 0.15\textwidth]{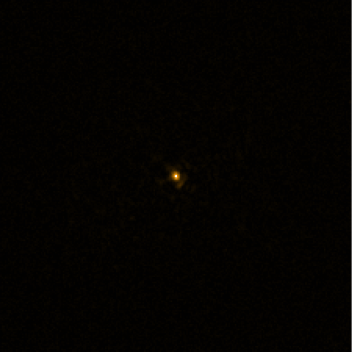}&
			\includegraphics[width= 0.15\textwidth]{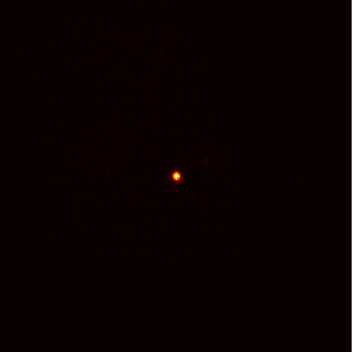}\\

			\footnotesize{(a) Initial speckle}&\footnotesize{(b) Wide illum.}&\footnotesize{(c) Speckle from }&\footnotesize{(d) Focused illum}&\footnotesize{(e) Focused illum}\vspace{-0.3cm}\\\\
			&\footnotesize{validation camera}&\footnotesize{single source}&\footnotesize{valid. camera}&\footnotesize{corrected emission}
			
		\end{tabular}
		\caption{Analyzing focusing inside tissue. each row visualizes a different experiment  on a different tissue slice. 
			(a)~Main camera when SLMs are blank, demonstrating initial speckles.
			(b)~Validation camera when illumination SLM is blank, demonstrating the enlightened bead layout at emission wavelength.
			(c)~Speckles from one bead (imaging SLM is blank and illumination SLM is corrected), demonstrating the amount of aberration.
			(d)~Validation camera when illumination is corrected, demonstrating that most light gets into a single spot (excitation wavelength).
			(e)~Main camera when both SLMs are corrected, demonstrating focusing in a single spot.
		}\label{fig:analyze}
	\end{center}
\end{figure*}

\begin{figure*}[t!]
	\begin{center}
		\begin{tabular}{@{}c@{~~}c@{~~}c@{~~}c@{~~}c}
			{\raisebox{1.1cm}{\rotatebox[origin=c]{90}{First iteration }}}&
			\includegraphics[width= 0.15\textwidth]{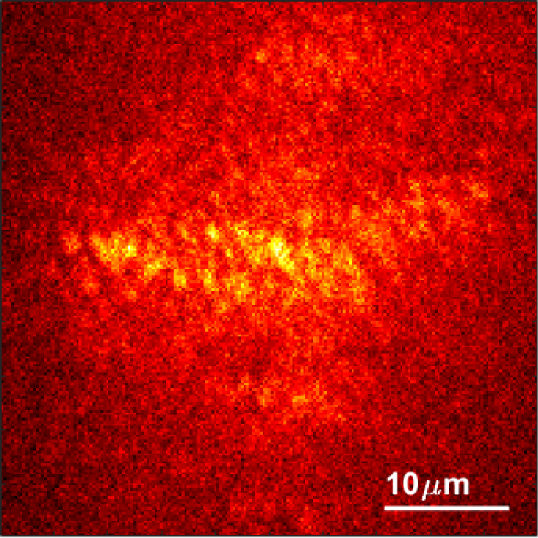}&
			\includegraphics[width= 0.15\textwidth]{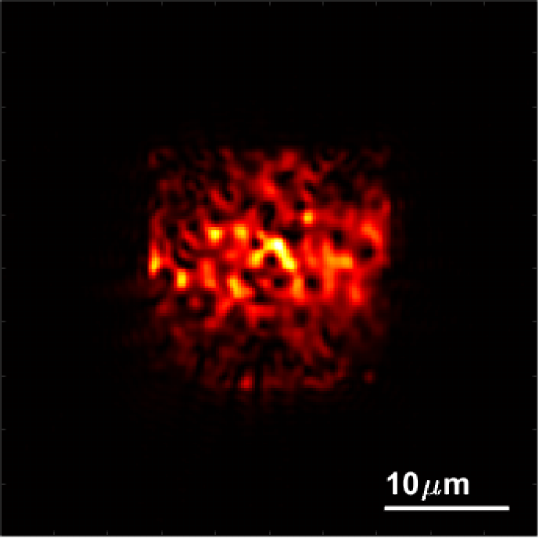}&
			\includegraphics[width= 0.15\textwidth]{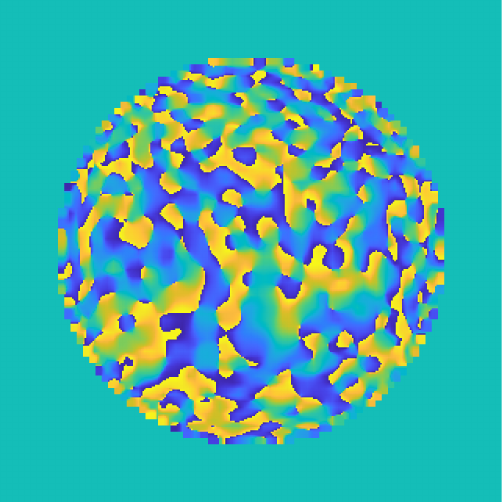}&
			\includegraphics[height= 0.15\textwidth]{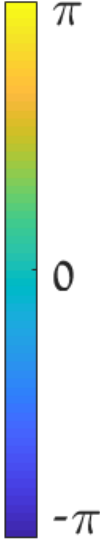}\\
			{\raisebox{1.1cm}{\rotatebox[origin=c]{90}{Last iteration  }}}&
			\includegraphics[width= 0.15\textwidth]{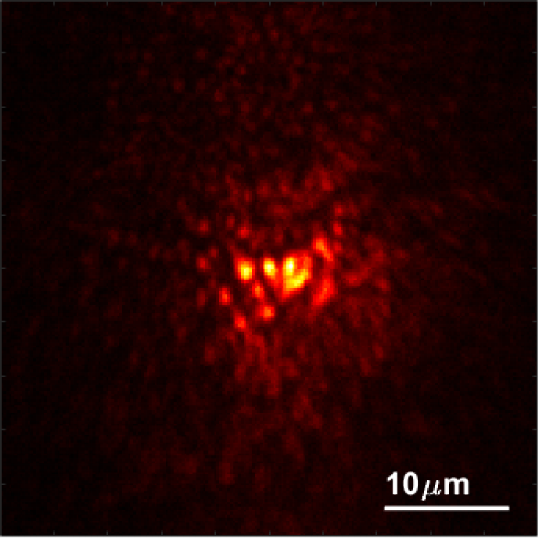}&
			\includegraphics[width= 0.15\textwidth]{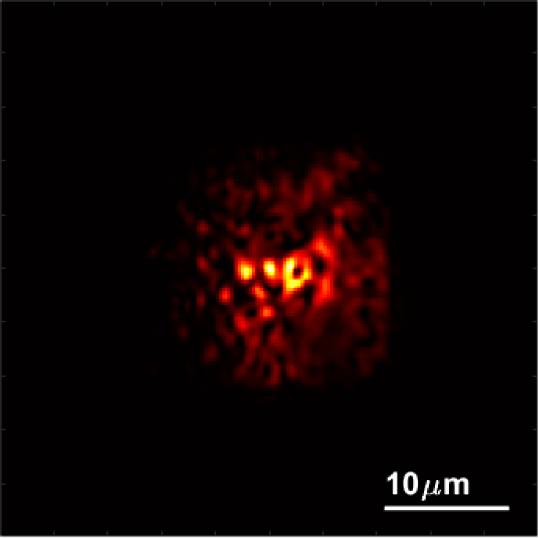}&
			\includegraphics[width= 0.15\textwidth]{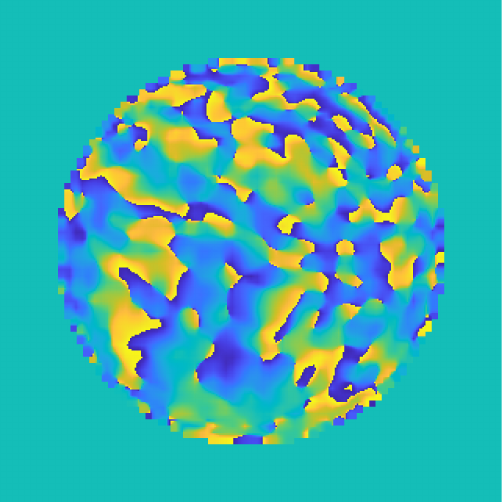}&
			\includegraphics[height= 0.15\textwidth]{figs/PR_part/colorbar.pdf}\\
			
			&\footnotesize{(a) Speckle image}&\footnotesize{(b) Recovered intensity} &\footnotesize{(c) Recovered phase}&\\
			& &\footnotesize{Sensor plane}&\footnotesize{Fourier plane}
		\end{tabular}
		\caption{Phase reconstruction: (a) Speckle image captured by the main camera in the first and last iteration of the algorithm. (b) The intensity of the recovered aberration correction in the sensor plane.  In the first iteration when multiple incoherent beads are excited we cannot fully explain the image as a single coherent wavefront. But as the algorithm converges to excite a single bead the recovered wavefront better matches with the captured image. (c) The phase of the recovered aberration correction in the Fourier plane, which is the mask presented on the SLM.
		}\label{fig:PR}
	\end{center}
\end{figure*}

\begin{figure*}[t!]
	\begin{center}
		\begin{tabular}{@{}c@{~~}c@{~~}c@{~~}c@{~~}c@{}}
			
			{\raisebox{1.6cm}{\rotatebox[origin=c]{90}{Reference}}}&
			\includegraphics[width= 0.22\textwidth]{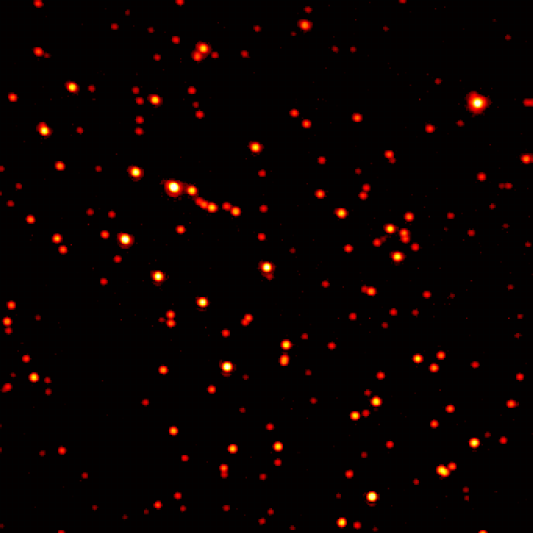} &	
			\includegraphics[width= 0.22\textwidth]{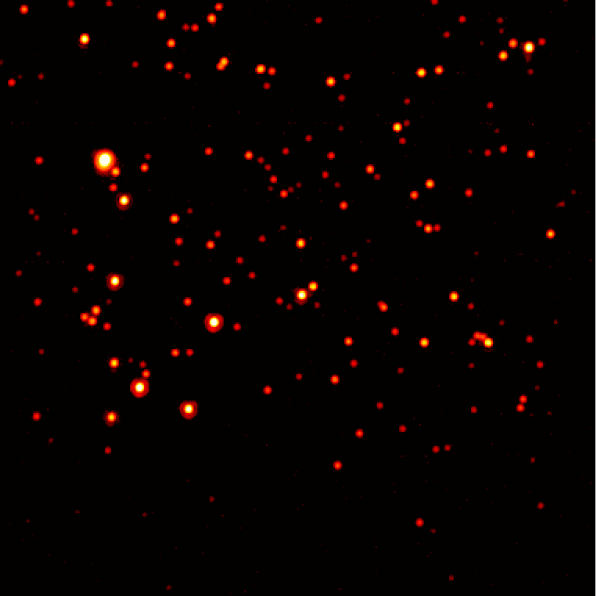} &
			\includegraphics[width= 0.22\textwidth]{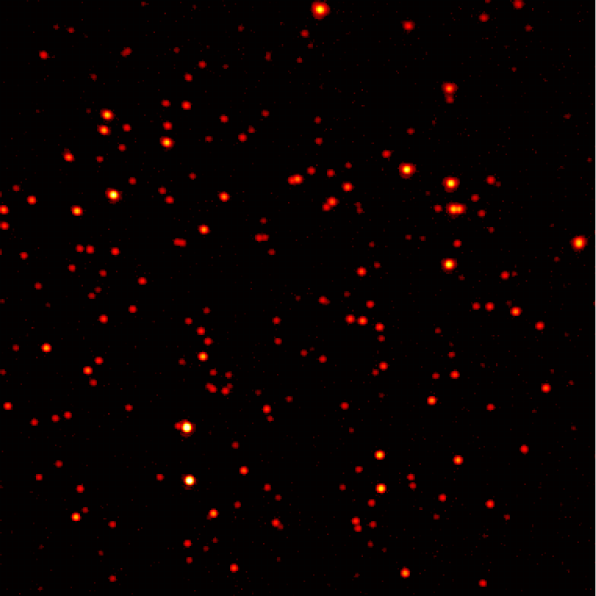} &		
			\includegraphics[width= 0.22\textwidth]{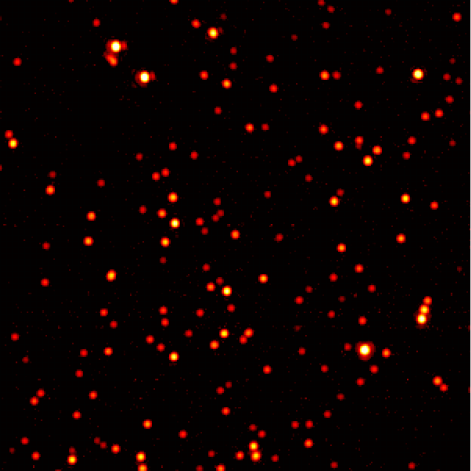} \\
			
			{\raisebox{1.6cm}{\rotatebox[origin=c]{90}{Simple ME}}}&
			\includegraphics[width= 0.22\textwidth]{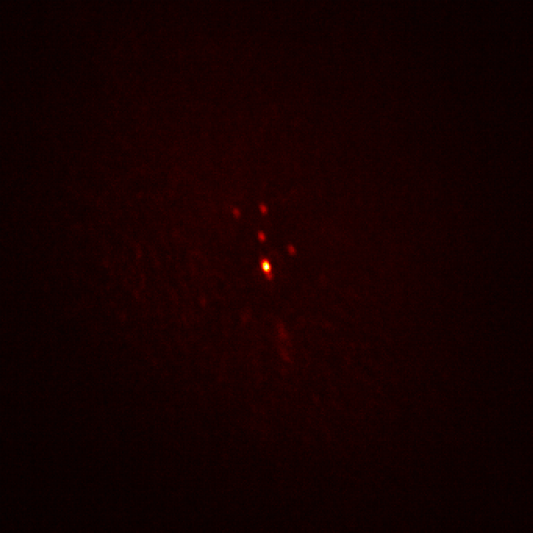} &	
			\includegraphics[width= 0.22\textwidth]{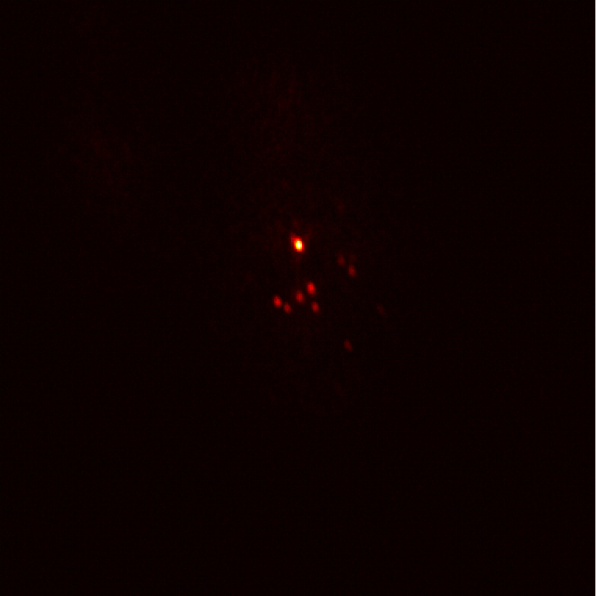} &
			\includegraphics[width= 0.22\textwidth]{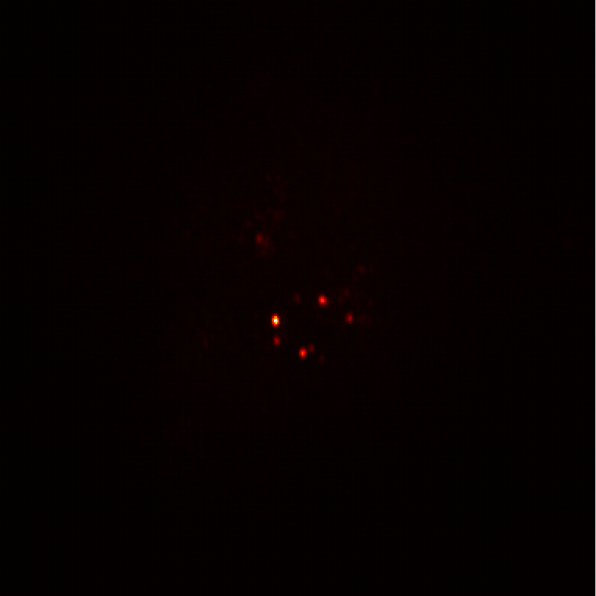} &				
			\includegraphics[width= 0.22\textwidth]{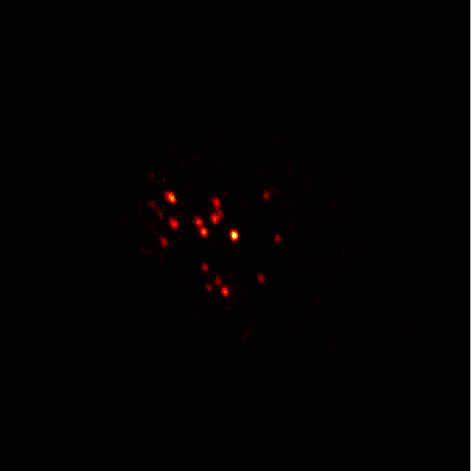} \\
			{\raisebox{1.6cm}{\rotatebox[origin=c]{90}{Tilt-shift ME}}}&
			\includegraphics[width= 0.22\textwidth]{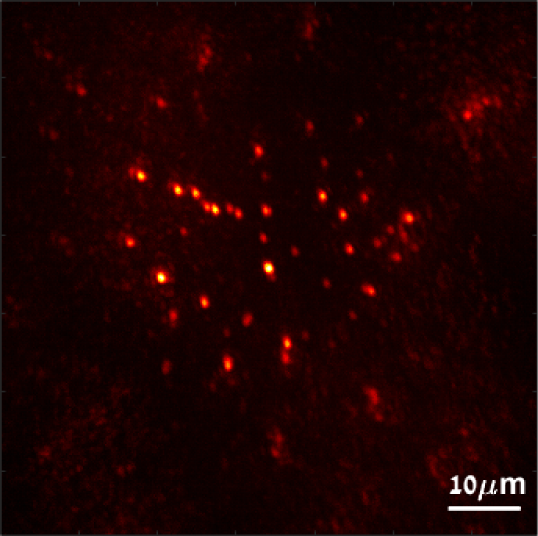} &	
			\includegraphics[width= 0.22\textwidth]{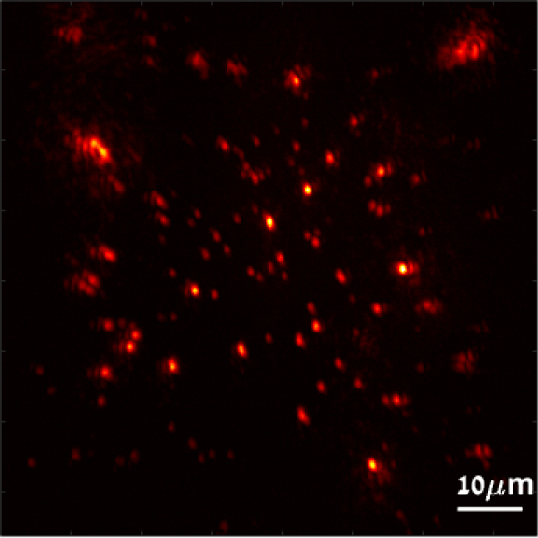} &
			\includegraphics[width= 0.22\textwidth]{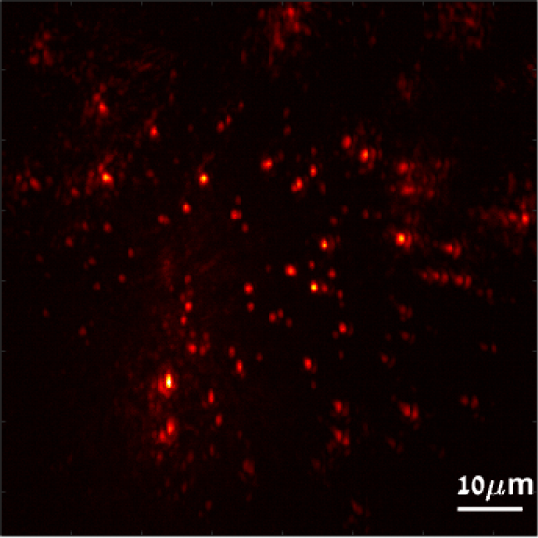} &	
			\includegraphics[width= 0.22\textwidth]{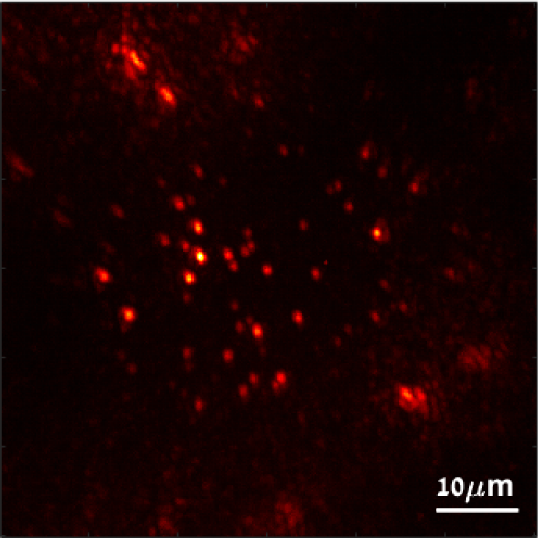} \\

		\end{tabular}
		\caption{			Imaging a wide area of fluorescent sources, behind 4 different tissue slices. Top: Reference image from validation camera. Second row: Imaging fluorescent sources within a small  region behind the tissue using a simplified memory effect. We  use the correction mask on the imaging arm, and use  wide illumination to excite multiple beads.  Third row: Using the tilt-shift memory effect to image a wider area, by shifting the imaging mask in the Fourier plane.	\Marina{in one of the previous versions you marked the relevant regions. It was easier to see}	}\label{fig:wide_area}
	\end{center}
\end{figure*}

	\section{Results}\label{sec:results}
	In our experimental implementation, we use fluorescent microspheres of diameter $\unit[200]{nm}$ (ThermoFisher FluoSpheres dark red), excited and imaged with $NA=0.5$ objectives so that the particles are slightly smaller than the diffraction limit. For excitation, we use a $\unit[637]{nm}$ laser, and to measure emission we use a band-pass filter of center wavelength $\unit[680]{nm}$ and bandwidth $\unit[10]{nm}$.
	\blue{In the main paper, we use  as scattering samples  chicken breast tissue slices of thickness $200-400\mu m$. In the supplement, we also show results using other scattering phantoms, including parafilm and polystyrene beads dispersed in agarose. 
		For all examples, significant scattering is present, and a standard microscope cannot image the actual source pattern. The beads are attached at the back of the tissue layer, separated only by a $150\mu m$ microscope cover glass.} We use two Pluto Holoeye SLMs, and a Prime BSI sCMOS sensor for imaging fluorescent emission. 
	\Marina{Do you need to have a section about the way you prepare the samples in the supplement?}

	\figref{fig:convergence} visualizes the power iterations of our algorithm from both the main camera and the validation camera. In the beginning the main camera sees a wide speckle pattern, and from the validation camera we can see that a wide speckle pattern reaches the back of the tissue. 
	We also use a band-pass filter on the validation camera to image the beads excited by each modulation pattern.
	The validation camera confirms that as the algorithm proceeds the illumination reaching the back of the tissue converges into a single spot. 
	Even if we manage to excite a single bead, the emitted light can scatter on its way to the main camera and generate a speckle pattern. 
	\blue{In  \figref{fig:convergence} we first visualize  this scattering  by showing the views of  the main camera  when modulation is used in the illumination SLM to focus the excitation, but with no modulation at the imaging arm.   In addition, we show what happens if the modulation pattern of each iteration is also placed on the SLM of the imaging arm.  }
	As the iterations proceed and the modulation pattern improves, the imaging SLM refocuses the light emitted from the excited bead  into a single sensor spot.

	When multiple fluorescing particles are present in the field of view, the algorithm typically converges to the strongest one. 
	However the particle at which the algorithm converges  may vary due to multiple reasons such as imaging noise, fluorescence bleaching or local minima of the phase diversity optimization. \blue{Convergence can also change if the optimization is initialized with a different speckle pattern.} 
	

	In \figref{fig:analyze} we demonstrate the final iteration of our algorithm  on a few additional examples.  
	In \figref{fig:analyze}(c) we also  visualize the actual scattering of the tissue layer.  
	To this end we place the correction mask on the illumination SLM only, bringing all light to excite a single bead. We use no correction on the imaging arm, allowing us to visualize the speckles from this  source.
	This image corresponds to a column of the transmission matrix $\TMo$.

	In \figref{fig:PR}, we visualize the phase diversity acquisition results at the first and last iterations of the algorithm. \figref{fig:PR}(a) shows the image from the main camera in both iterations when the imaging SLM applies no correction. 
	\figref{fig:PR}(b) shows the intensity of the recovered wavefront at the plane of the camera sensor, which is set conjugate to the plane of the fluorescent sources. The phase diversity optimization attempts to explain the images in  \figref{fig:PR}(a). However, in the first iteration, the captured speckle image is an incoherent summation from multiple particles, which the optimization objective of \equref{eq:phase-diversity-opt-main} attempts to explain with a single coherent wavefront; thus the result is imperfect. In the last iteration, the algorithm  excites a single particle, and indeed the estimated wavefront better explains the captured intensity. 
	Finally, \figref{fig:PR}(c) shows the retrieved phase in the frequency domain, which is essentially the pattern displayed on the SLM. We note that, as we use a phase-only SLM, we effectively correct only the phase of the wavefront and neglect its amplitude.

	\paragraph{Imaging a wide field of view.} 
	The recovered modulation pattern is designed to focus at a single particle inside the tissue sample. However, due to the memory effect, the corrections of nearby spots are similar. We demonstrate this experimentally in the second row of \figref{fig:wide_area}: We place a random pattern on the illumination arm, which results in exciting multiple fluorescing particles. We then place the correction pattern recovered for focusing on one of the fluorescing particles on the imaging arm. We observe that, thanks to the memory effect, the camera can image a small neighborhood of particles around the focus points, and not just the single particle the correction pattern corresponds to. To further improve on this, we use the tilt-shift memory effect~\cite{osnabrugge2017generalized,single-sct-iccp-21} and shift the modulation mask in the Fourier plane. As we explain in the supplement, this shift allows us to focus at nearby regions using the same correction pattern. In the third row of \figref{fig:wide_area}, we image a wider range of particles behind the tissue sample, by scanning $21\times 21$ such shifts. 
	We acknowledge that by placing the SLM at a plane conjugate to the sample itself~\cite{Mertz:15,osnabrugge2017generalized} rather than in the Fourier plane, we can probably expand the region corrected by a single modulation and reduce the number of required shifts.
	Even after exploiting the tilt-shift, the extent of the memory effect is limited, and beads at the periphery of the images of the last row in \figref{fig:wide_area} are either not recovered or strongly aberrated. Imaging beyond this region  would require applying another set of power iterations to calibrate a different wavefront modulation, amplifying the importance of the faster convergence of our proposed procedure.
	\section{Discussion}
	
	We extended iterative phase conjugation algorithms to apply to incoherent fluorescent imaging. Even though the incoherent contribution of different sources alters the linear transmission model from which these algorithms are derived, we show that the non-linear incoherent model accelerates convergence rate, from exponential to doubly-exponential. To find a wavefront correction pattern, we need to excite the tissue with a very small number of trial patterns and measure the resulting excitation. The number of measurements is orders of magnitude smaller than that of previous optimization-based techniques.

	We used the recovered modulation pattern to image fluorescent particles placed behind a tissue sample. However, wavefront correction in thick tissue is spatially-varying, and each modulation pattern is only usable for imaging a limited region, with size determined by the memory effect. To image a wider region behind the tissue sample, we have to apply the modulation recovery algorithm multiple times in different sub-regions. This makes it even more important to have fast wavefront shaping    algorithms.  
	One way to further reduce the number of acquired images is to use a tilt-shift adaptation of the aberration correction estimated in one region, and initialize with it the power iteration in neighboring sub-regions.


	
	Our current results apply only on a sparse set of fluorescent particles. 
	Increasing the density of the sources is challenging  because as speckle contrast decays~\cite{SeeThroughSubmission},
	it is harder for the phase diversity acquisition scheme to recover  phase. To alleviate this problem, we could adopt other phase acquisition schemes, such as using a Shack-Hartmann sensor~\cite{Shack71}. 
	
	\blue{Another issue which may challenge convergence with a dense continuous fluorescent object is two nearby spots emitting similar power.
		This is due to the fact that when a transmission matrix contains multiple eigenvectors with the same eigenvalue, power iterations may not separate them, and can converge to a linear combination of the two eigenvectors. 
		We note however that the incoherent convergence rate as analyzed in supplement Eq. (26) depends not only on the actual eigenvalues, but also on the initial excitation $\rkTMi\bu^{(0)}$. As this excitation is usually a highly varying speckle pattern, there is a better chance to separate between nearby illuminators of similar power.} 
	
	Our approach also relies on the assumption that the  excitation and emission wavelength are close enough so that the excitation and emission transmission matrices  are sufficiently similar. \blue{Also, as the emitted light contains  multiple  wavelengths, these can produce somewhat different speckle patterns.  } There is evidence in the literature that speckle patterns produced by nearby wavelengths are correlated~\cite{Zhu:20}, but this similarity degrades as the tissue sample thickness increases\blue{~\cite{Andreoli_2015}}. In our experimental implementation, there is a $\unit[40]{nm}$ gap between the emission and excitation wavelengths. In linear fluorescence imaging, the gap between excitation to emission can be made  lower than that, leading to even stronger correlation between the wavefronts. \Marina{the convergence depends on the sample thickness in some way? or beads density? Or only the phase recovery?}

	\paragraph{Relationship to memory-effect techniques.} Our work is orthogonal to approaches for imaging fluorescent sources through tissue using speckle statistics, and in particular the memory effect~\cite{Bertolotti2012,Katz2014,SeeThroughSubmission}.\Marina{do you want to cite ours also?} Recently, such approaches were successful in imaging a sparse set of fluorescent particles inside tissue, using hundreds~\cite{Boniface2020,Zhu22} or even just dozens~\cite{WeiYu22} of images. While the field of view of a wavefront shaping modulation is constrained by the extent of the memory effect, as demonstrated in \figref{fig:wide_area},   approaches based on the memory effect can recover full-frame patterns with much wider field of view. By contrast, memory effect correlations only exist in thin tissue layers, while approaches based on phase conjugation can theoretically achieve larger penetration depths. However, in practice, the penetration depth is greatly constrained by the very weak signal-to-noise ratio of fluorescent emission. Approaches based on phase conjugation make it possible to
	not only \emph{image} through scattering, but also \emph{focus light}  inside scattering tissue, a capability that memory effect approaches lack. Focusing inside tissue is important for applications such as laser treatment therapy, confocal microscopy, and STED microscopy.
	Finally, a modulation recovered from fluorescent sources can also be used  to image adjacent non-fluorescent tissue structures. 
	\subsection*{Acknowledgments} We thank Lucien Weiss and Onit Alalouf for their help preparing experimental data.
	\subsection*{Disclosure} The authors declare no conflicts of interest.

\section{Supplementary}
\subsection{Convergence analysis for incoherent transmission}\label{sec:proof}
We analyze the convergence of our iterative algorithm, while modeling the incoherent summation of different fluorescent emitters. We show that incoherence leads to asymptotically faster  convergence when compared to the coherent case.   

\paragraph{Model.}
We introduce some notation that we will use to rewrite the transmission image formation model in an equivalent form that is more convenient for our analysis. For this, we recall from the main paper our assumption that the transmission matrices for excitation $\TMi$ and emission $\TMo$ are transposes of one another, $\TMo={\TMi}^\top$. Therefore, the rows of $\TMi$ equal the columns of $\TMo$. We write $\rkTMi,\ckTMo$ for the $k$-th row and column of these $\TMi$ and $\TMo$, respectively. We also denote by $n_k$ the norm of the $k$-th row of $\TMi$ and $k$-th column of $\TMo$,
\BE
\blue{n_k\equiv\sum_x |\TMi_{k,x}|^2=\sum_x |\TMo_{x,k}|^2,}
\EE
where $x$ is a position on the input or sensor plane, for excitation and emission respectively. $n_k$ can be lower than one, because some light emitted by the fluorescent particles scatters at angles higher than the numerical aperture of the objective and does not reach the sensor. We use $\nTMi,\nTMo$ to denote the matrices  $\TMi$,  $\TMo$ after normalizing their rows and columns, respectively, to have unit-norm:
\BE
\nrkTMi\equiv\frac{1}{\sqrt{n_k}}\rkTMi,\quad \nckTMo\equiv\frac{1}{\sqrt{n_k}}\ckTMo.
\EE 
We will account for the norm explicitly in the image formation model. Then, linear fluorescence emission from the $k$-th location inside tissue is proportional to 
\BE \label{eq:v-with-e-n} 
|\bou_k|^2=e_k n_k |\nrkTMi \bu|^2,
\EE  
where we use $e_k$ to denote the emission power of the $k$-th particle \blue{(for simplicity of  exposition, in the main  text we have absorbed $e_k$ into the unnormalized transmission matrix)}. Similarly, the measured emission intensity is 
\BE\label{eq:int-with-e-n}
\sum_k |\ckTMo|^2|\bou_k|^2=\sum_k |\nckTMo|^2n_k |\bou_k|^2.
\EE

With this notation we can express the combined transmission operator as 
%
%
%
%
\BE\label{eq:Ta-diag}
\TMa \equiv \nTMo \bW \nTMi,
\EE
where $\bW$ is a diagonal matrix with non-negative diagonal entries 
\BE\label{eq:wk-def} 
w_k\equiv n_k\sqrt{e_k}.
\EE 
These entries encode the power of the fluorescent emitter at the $k$-th focus location, as well as the amount of energy transferred on the $k$-th row and column of $\TMi$ and $\TMo$, respectively. 


To further simplify our analysis, we also assume that the wavefonts emitted by different fluorescent particles are sufficiently random, and their correlation
\BE\label{eq:decorr}
\blue{\eps_{k,\ell}\equiv\sum_x \nTMo_{x,k}{\nTMo_{x,\ell}}^*}
\EE
\blue{is sufficiently small. For the rest of this derivation we will assume $\eps_{k,\ell}\approx 0$ and can be neglected. We note that the memory effect implies that the rows of the transmission matrix are correlated {\em shifted} versions of each other; that is $\nTMo_{x,k}\approx \nTMo_{x+\Dl,\ell}$ where $\Dl$ is the displacment between the $k,\ell$ particles.
	However, even in the presence of ME correlation, at the zero shift we consider in \equref{eq:decorr}, such rows are uncorrelated, as effectively, the row entries   are pseudo-random  patterns. }

With this decorrelation assumption, $\nTMi,\nTMo$ are orthogonal matrices, and the diagonal entries $w_k$ of the matrix $\bW$ in \equref{eq:Ta-diag} will correspond to the eigenvalues of the transmission operator $\TMa$.

\paragraph{Power method under coherent illumination.}
We start by reviewing the principles of the power method~\cite{trefethen97} considering a coherent illumination model. For simplicity, we assume that we can measure the phase of wavefronts rather than only their intensities. We will later extend this analysis to the incoherent case.

To apply the power method for the coherent illumination case, we start with a random illumination pattern $\bu^0$. At each iteration, we illuminate the tissue sample with a wavefront $\bu^t$. The propagation through the sample and back to the sensor produces a wavefront $\TMa \bu^t$. Assuming we can measure both amplitude and phase of the resulting wavefront, we
update the illumination wavefront as 
\BE\label{eq:norm-utp1}
\bu^{(t+1)}\equiv\frac{(\TMa \bu^{(t)})^*}{\|\TMa \bu^{(t)}\|},
\EE
where we normalize  $\bu^{(t+1)}$ to fix the total energy of the excitation pattern $\|\bu^{(t+1)}\|$ at each iteration, as determined by the power of the excitation laser. We note that, in practice, we use a phase-only SLM, and thus we only display the phase of $\bu^{(t+1)}$, dropping its amplitude.

To understand the convergence of this algorithm, we denote by $\beta_k^t$ the energy scattered from the $k$-th particle  inside the tissue sample at the $t$-th iteration, 
\BE
\beta_k^{(t)}\equiv w_k \nrkTMi \bu^{(t)}.
\EE
With this notation, the operation of the transmission matrix $\TMa$ on $\bu^{(t)}$ equals the sum of the columns of $\nTMo$ weighted by $\beta_k^{(t)}$,
\BE
\TMa \bu^{(t)}= \sum_k \beta_k^{(t)} \nckTMo
\EE
Therefore, the illumination pattern used at the next iteration will equal a weighted combination of the conjugate columns,
\BE\label{eq:u_tp_1}
\bu^{(t+1)}=\left(\TMa \bu^{(t)}\right)^*=\sum_k \beta_k^{(t)}  {\nckTMo}^*,
\EE
$\bu^{(t+1)}$ is then normalized as in \equref{eq:norm-utp1}.

Applying $\nTMi$  on $\bu^{(t+1)}$ can be expressed as a summation over all entries $x$.
Using the decorrelation assumption of \equref{eq:decorr}, and \equref{eq:u_tp_1} this reduces to
\BE
\blue{\nrkTMi \bu^{(t+1)}\!=\!\sum_x \!\nTMi_{k,x} \bu^{(t+1)}_x\!=\! \sum_{\ell}\beta_{\ell}^{(t)}\! \sum_x \!\nTMi_{k,x} {\nTMo_{x,l}}^*\! =\! \beta_k^{(t)}.}
\EE
Therefore,
\BE
\beta_k^{(t+1)}=w_k \nrkTMi \bu^{(t+1)}=w_k \beta_k^{(t)}.
\EE
A simple recursion implies that $\beta_k^{(t)}$ follows an exponential series of the form
\BE\label{eq:exp-eigv-coherent}
\beta_k^{(t)}=(\lambda_k)^{t} c_k
\EE
where $\lambda_k\equiv w_k, c_k\equiv \beta_k^0$.
\equref{eq:exp-eigv-coherent} states that the entries of $\beta^{(t)}$ scale exponentially as a function of the iteration number. This implies that the gap between the largest entry of $\beta$ and the next one increases with each iteration. It is easy to show that the sequence converges quickly into a one-hot vector, which is non-zero at a single entry.

\paragraph{Power method under incoherent illumination.}
The coherent case is similar to the classical application of the power method. We now make the necessary adaptations for the incoherent case. 
The convergence rate we  achieve is asymptotically faster than the exponential convergence we derived in \equref{eq:exp-eigv-coherent}.
Throughout this derivation we assume the power of the fluorescent sources is constant during optimization and ignore effects such as blinking or bleaching.  

To study the incoherent emission of fluorescent sources, we start by deriving the corresponding image formation model. At the $t$-th iteration, we excite the tissue sample with a wavefront $\bu$, and measure
\BE
I^{(t)}=\sum_k |\nckTMo|^2 \alpha_k^{(t)},
\EE
where $\alpha_k^{(t)}$ denotes the incoherent equivalent of $\beta_k^{(t)}$, the energy emerging from the $k$-th emitter, times the norm of the $k$-th column. 
Following the definitions in Eqs. (\ref{eq:v-with-e-n}), (\ref{eq:int-with-e-n}) and (\ref{eq:wk-def}),
we use:
\BE\label{eq:alpha-k}
\alpha_k^{(t)}=n_k |\bou_k|^2=w_k^2 |\nrkTMi \bu^{(t)}|^2.
\EE
Our goal is to show that, as in the coherent case, within a small number of iterations, $\alpha_k^{(t)}$ converges to a one-hot vector. 

At each iteration, our algorithm needs to estimate some phase from the speckle intensity image $I^{(t)}$. As we mention in the main paper, we use a phase diversity acquisition scheme. As this scheme is based on optimizing a non-linear score, analyzing its convergence is not straightforward.
To this end, we start by considering a simpler acquisition scheme based on point diffraction interferometry~\cite{smartt1975theory, akondi2014digital}. 
This scheme is highly-sensitive to noise, and implementing it using weak fluorescent sources is impractical. However, the advantage of this scheme is in providing a closed-form expression for the contribution of different sources to the estimated phase, allowing for simple analysis. We will later use numerical simulations to compare the convergence of phase diversity  against the analytical expressions we derive from point diffraction interferometry.

\paragraph{Point diffraction interferometry.} 
Consider the field $\nTMo_{x,k}$ generated by the $k$-th fluorescent source at image point $x$. We decompose it as \BE \nTMo_{x,k}=\nvTMo_{x,k}+\mTMo_k\EE
where $\mTMo_k$ is its complex mean
\BE
\blue{\mTMo_k=\sum_x \nTMo_{x,k}} 
\EE\Anat{need normalization?}
and $\nvTMo_{x,k}=\nTMo_{x,k}-\mTMo_k$.

Point diffraction interferometry captures $J\geq 3$ images using the SLM in the Fourier plane of the imaging arm. It changes phases at a single spot corresponding to the  $0$-th (central) frequency. 
Placing phase $\phi_j$ at the zero frequency only changes the mean of the signal, and the intensity to be measured at pixel $x$ of the image plane from the $k$'th source corresponds to 
\BEA\label{eq:I-inc-weighted-sum}
I_{k,x}^{(t,j)}&=&\alpha_k^{(t)} \left|\nvTMo_{x,k}+e^{i\phi_j}\mTMo_k\right|^2\\&=&\alpha_k^{(t)}\left(\left|\nvTMo_{x,k}\right|^2+\left|\mTMo_k\right|^2+2\Re{\left(e^{i\phi_j}\mTMo_k {\nvTMo_{x,k}}^*\right)}\right),\nonumber
\EEA
where $\Re$ denotes the real component, and $\alpha_k^{(t)}$ defined in \equref{eq:alpha-k} corresponds to the energy emitted by the $k$-th fluorescent particle given  the current excitation wavefront. In the presence of multiple incoherent sources, we measure the incoherent summation of the intensity speckle patterns produced by each of them,
\BE
I^{(t,j)}_x=\sum_{k} \alpha_k^{(t)} I^{(t,j)}_{k,x}
\EE


In point diffraction interferometry, we capture $J\geq3$ images with equally-spaced phase shifts $\phi_j=[1\ldots J]\frac{2\pi}{J}$.
Standard phase shifting interferometry techniques~\cite{hariharan1987digital} imply that, by summing measurements with different phase shifts, we can extract
\BE
\bu^{(t+1)}_x=\sum_j e^{-i\phi_j} I^{(t,j)}_x =\sum_k \alpha_k^{(t)} \mTMo_k\nvTMo_{x,k}^*.
\EE
Thus, point diffraction interferometry extracts a weighted combination of the wavefronts emerging from all incoherent sources. The weights correspond to the intensity they receive from the previous excitation pattern, weighted by a complex scalar corresponding to the mean $\mTMo_k$.

In the next iteration of the power method, we excite the tissue with the extracted wavefront normalized to have unit energy:
\BE\label{eq:normalize}
\frac{{\bu^{t+1}}}{\|\bu^{t+1}\|}
\EE


\paragraph{Convergence.}
We now show that the sequence $\alpha_k^t$ converges to a one-hot vector, which implies that the iterative approach we described above converges.
To this end, we note that when we excite the tissue sample with illumination ${\bu^{t+1}(x)}$, we effectively multiply ${\bu^{t+1}(x)}$ by $\nTMi$.
Using the decorrelation assumption in \equref{eq:decorr}, we can write the energy at the $k$-th fluorescent particle as
\BE
\nrkTMi {\bu^{(t+1)}_x}=\sum_\ell \alpha_\ell^{(t)} \mTMo_\ell \sum_x \nTMi_{k,x} {\nvTMo}_{x,\ell}^*= \alpha_k^{(t)}  \mTMo_k.
\EE
Thus, using the definition of $\alpha_k$ in \equref{eq:alpha-k} and ignoring the normalization in \equref{eq:normalize}, we have
\BE
\alpha_k^{(t+1)}=(w_k \alpha_k^{(t)}  |\mTMo_k|)^2.
\EE
Using recursion, this leads to
\BEA\label{eq:exp-decay}
\alpha_k^{(t)}&=&(w_k)^{2(2^{t+1}-1)}  |\mTMo_k|^{2(2^{t}-1)}  |\nrkTMi \bu^{(0)}|^{2^{t+1}}\nonumber\\
&=&(\lambda_k)^{2^t}\cdot c_k. 
\EEA
where 
\BE
\lambda_k=w_k^4 |\mTMo_k|^2 |\nrkTMi \bu^{(0)}|^2, \quad c_k=w_k^{-2} |\mTMo_k|^{-2}.
\EE
To understand the difference between this result and the coherent case in  \equref{eq:exp-eigv-coherent}, we note that in the coherent case the leading term converges as $\lambda_k^t$, which is an exponential sequence. In the incoherent case we have another exponent, and the leading term in \equref{eq:exp-decay} is of the form $(\lambda_k)^{2^t} $. This is known as a {\em doubly exponential series}, which will converge into a one-hot vector much faster than the exponential series of \equref{eq:exp-eigv-coherent}. 
%

\paragraph{Phase diversity acquisition.}
As we mentioned above, the point diffraction interferometry scheme is useful for analysis, as it leads to closed-form expressions. In practice, this approach is very sensitive to noise, and implementing it with weak fluorescent sources is unrealistic. Instead, our implementation uses a phase diversity acquisition scheme~\cite{MUGNIER20061}. We place $J=5$ known modulation patterns $H_j$ on the SLM of the imaging arm, and measure speckle intensity images of the form
\BE
I^{(t,j)}=\sum_k |h^j\star \nckTMo |^2 \alpha_k^{(t)},
\EE
where $\star$ denotes convolution, and $h^j$ is the Fourier transform of the pattern we placed on the SLM. We use gradient descent optimization to find a complex wavefront $\bu^{(t+1)}$ minimizing
\BE\label{eq:phase-diversity-opt}
\sum_j \left|I^{(t,j)}-|h^j\star \bu^{(t+1)}|^2\right|^2.
\EE
This optimization is subject to local minima, and it is hard to give any analytic guarantees about its convergence.
Below we conduct numerical simulations comparing its empirical convergence to the analytical predictions from point diffraction interferometry.

\begin{figure}[t!]
	\begin{center}
		
		\begin{tabular}{@{}c@{~~}c@{}}
			\includegraphics[width= 0.22\textwidth]{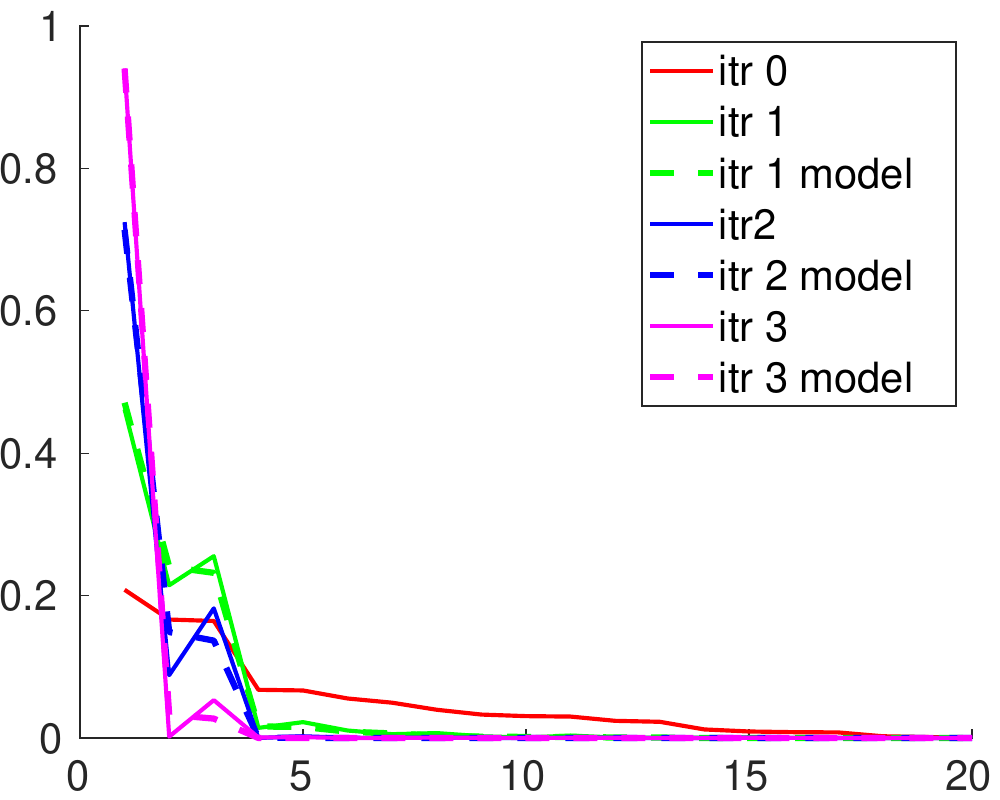}&
			\includegraphics[width= 0.22\textwidth]{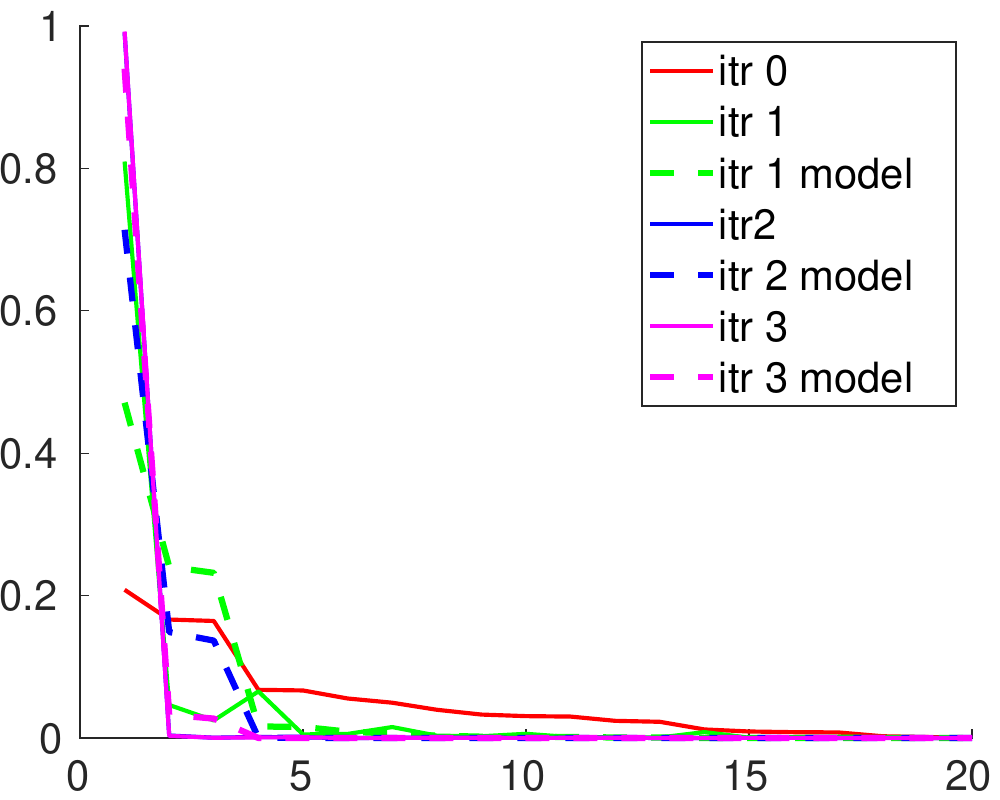}\\
			\includegraphics[width= 0.22\textwidth]{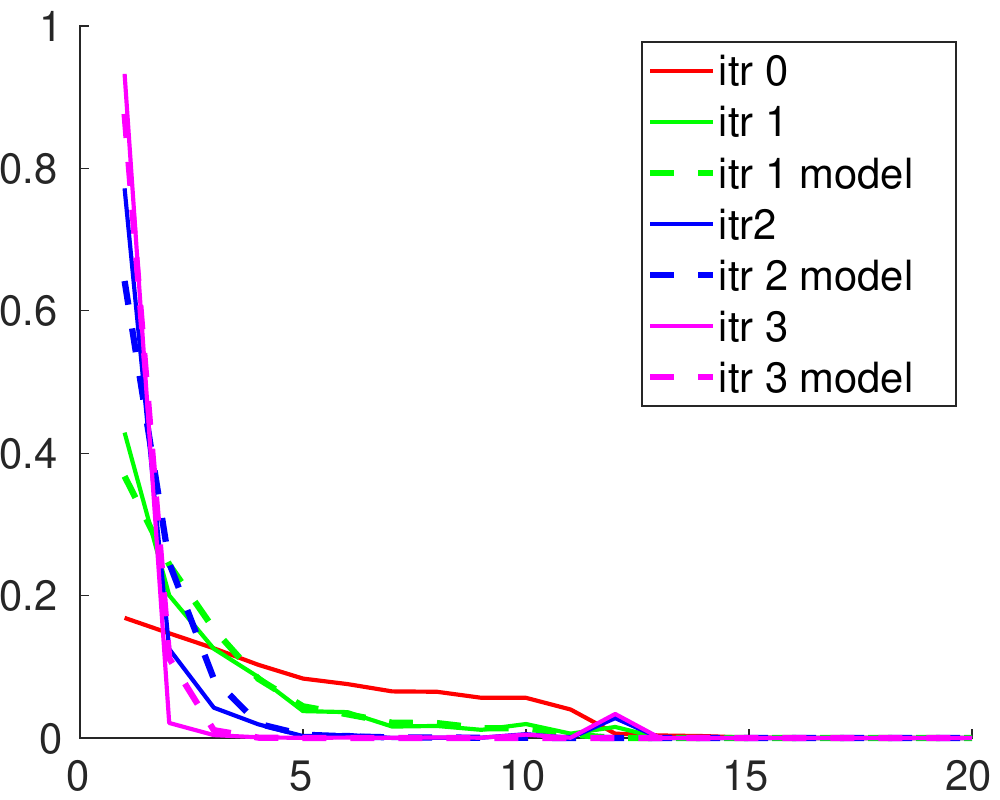}&
			\includegraphics[width= 0.22\textwidth]{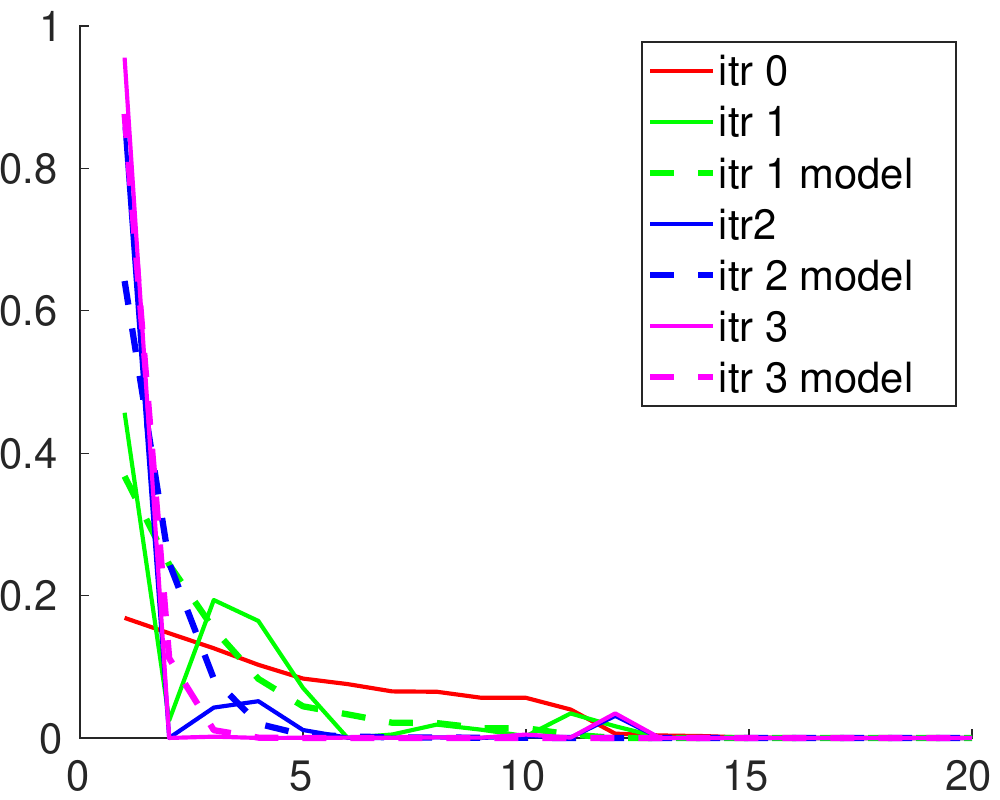}\\
			(a) Point diffraction interf.&(b) Phase diversity\\
			\multicolumn{2}{c}{	\includegraphics[width= 0.42\textwidth]{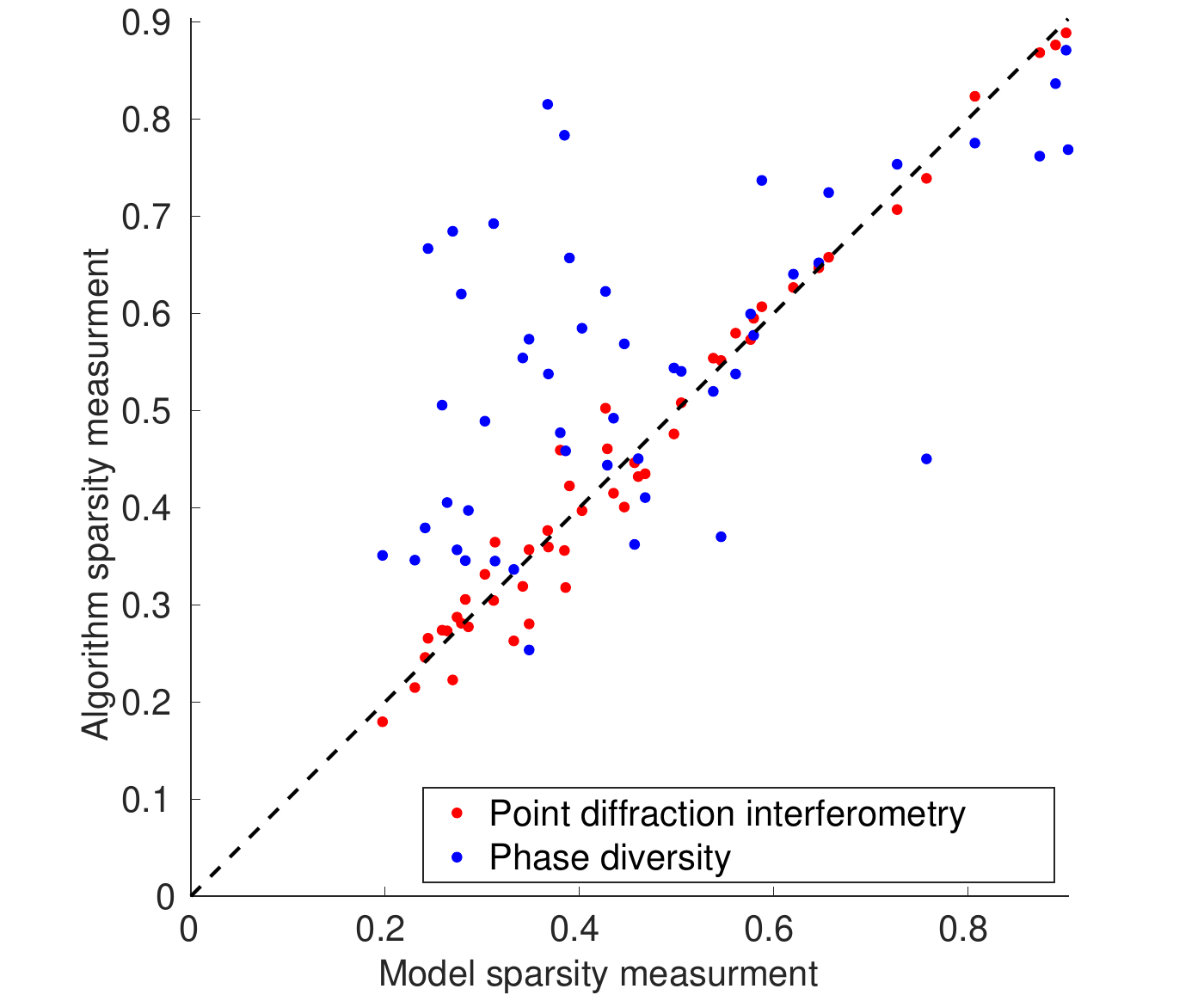}}\\
			\multicolumn{2}{c}{	(c) Sparsity statistics}
		\end{tabular}
		\caption{Numerical convergence evaluation. (a-b) two typical examples for the $\alpha^{(t)}$ values corresponding to the energy in different fluorescent sources for the first three iterations of the algorithm. We compare the convergence with an idealized noise free point diffraction interferometry scheme and the phase diversity optimization we use in practice. In each case we plot in dashed lines the prediction of the model in \equref{eq:exp-decay}, with qualitative match to what we measure in practice. The graphs are plotted as a function of the bead index $k$, where for ease of visualization we sort the eigenvalues in decreasing order of the bead strength $w_k$ so that the strongest eigenvalue appears in the first place.  (c) Comparing the sparsity of $\alpha^{(1)}$ to the model prediction for 50 random transmission matrices. Examples that match the model prediction should lie on the dashed diagonal line and indeed point diffraction interferomety results are concentrated around the diagonal line. Phase diversity optimization is often sparser than the model prediction (above the diagonal line) but can also be of lower quality. \Marina{possible to make everything larger? Why do you need to compare point diffraction interf. to model? Isn't this the same model? Just to show that they do not match because of sampling? Why is it useful?}	}\label{fig:simulation}
		\vspace{-0.4cm}
	\end{center}
\end{figure}

\paragraph{Numerical evaluation.}
We sample transmission matrices $\nTMo,\nTMi$ such that each row is a random i.i.d. complex Gaussian random vector. For simplicity all rows have the same mean $\mTMo_k$. We transform the noise vectors to the Fourier domain and set to zero any frequency above $NA=0.5$. In the primal domain, we limit the speckles in a Gaussian window of STD $20{\mu m}$, as the speckles imaged in our setup have a limited support and do not spread over the full sensor. We assume $K=20$ fluorescent sources. 
We initialize with a uniform excitation 
and apply the power method as we described above. We simulate phase acquisition with ideal noise-free point diffraction interferometry, and also by solving the phase diversity optimization of \equref{eq:phase-diversity-opt}, which may converge to local optima.  
In \figref{fig:simulation}(a-b) we plot the vectors $\alpha^{(t)}$ we obtain in the first three\Marina{four you mean incouding itr 0? In the plot caption it says three iterations} iterations of the algorithm. For ease of visualization, we sort the entries of this vector by decreasing order of $w_k$, so that the maximal eigenvalue is always at $k=1$. We also normalize the plotted vectors to sum to $1$. With both acquisition schemes, within a small number of iterations $\alpha^{(t)}$ is a one-hot vector.  
In each case we plot a dashed line with the expected values following the model of \equref{eq:exp-decay}. Even the point diffraction interferometry values do not  match this model precisely,  because the transmission matrices we sample have random rows which have low correlation, yet their correlation is not precisely zero as assumed in \equref{eq:decorr}. Despite this difference, the convergence rate of both schemes qualitatively agrees with the model predictions.

To statistically assess the differences between the model of \equref{eq:exp-decay} and the empirical phase retrieval results, we sample $50$ random transmission matrices and apply on each  the first iteration of the power method. 
For this, we use an initial excitation such that $|\nrkTMi \bu^{(0)}|$ is uniform.
We measure intensities using the point diffraction interferometry or phase diversity schemes, recover the phase of $\bu^{(1)}$, and compute the vector
\BE
\alpha^{(1)}_k=w_k^2|\nrkTMi\bu^{(1)}|^2.
\EE
To assess the sparsity of this vector, we measure
\BE
s=\max_k \frac{\alpha^{(1)}_k}{\sum_k \alpha^{(1)}_k}.
\EE
Ideally we want $s$ to be as close to $1$ as possible.
According to the model in \equref{eq:exp-decay} the sparsity of the eigenvalues after one iteration should be equivalent to
\BE
s_o=max_k \frac{w_k^6}{\sum_k w_k^6}.
\EE

In \figref{fig:simulation}(c), we evaluate $s_{pdi}$ and $s_{pd}$ using the point diffraction interferometry and phase diversity schemes for $50$ different transmission matrices.
For the $k$-th random transmission matrix, we plot the 2D points $(s^k_o,s^k_{pdi})$ and $(s^k_o,s^k_{pd})$.
The plot demonstrates that, in practice, the sparsity of phase diversity is equivalent or even better than point diffraction interferometry.
If $s_{pdi}$ and $s_{pd}$ matched the $s_o$ prediction,  all points should lie on the diagonal dashed line marked in the figure. We see that, for point diffraction interferometry, $s_{pdi}$ is proportional to  $s_o$, but is not exactly equivalent to it, as the decorrelation assumption of \equref{eq:decorr} does not hold exactly.
We obtained the result of phase diversity acquisition using gradient descent optimization, which does not always converge to a global optimum. 
The plot in \figref{fig:simulation}(c) illustrates that, 
in most cases, this solution is actually better than the  $s_o$ prediction (points above the dashed line),
though  for some transmission matrices the solution is worse, and the points lie below the dashed lines. 
\Marina{I didn't get what is the main message from this plot. That phase diversity can produce bad results but still when it produces good results, they are better? }
\begin{figure}[t!]
	\begin{center}
		\begin{tabular}{c}
			\begin{tabular}{ccc}
				\includegraphics[width= 0.13\textwidth]{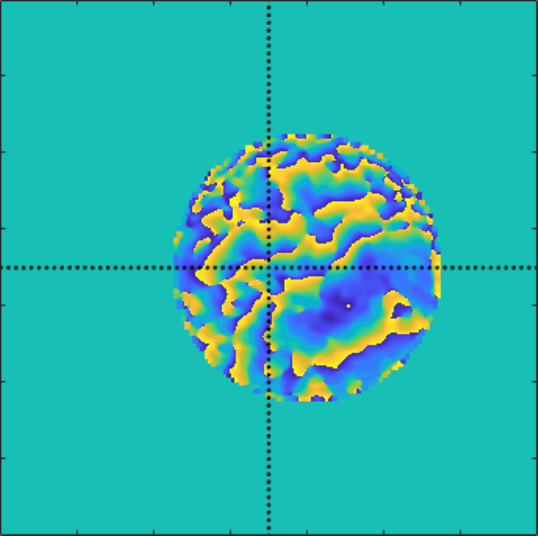} &
				\includegraphics[width= 0.13\textwidth]{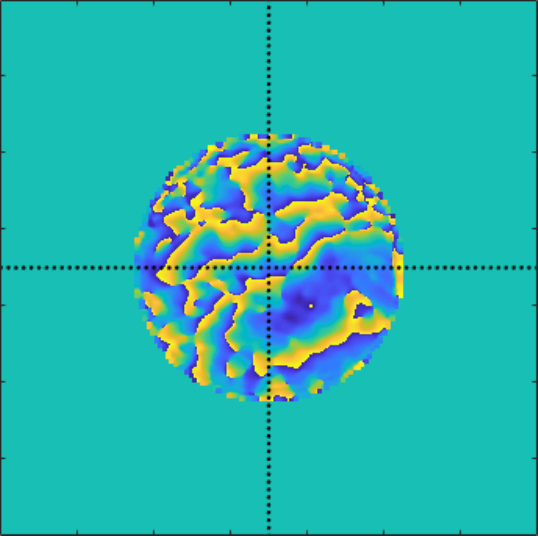} &
				\includegraphics[width= 0.13\textwidth]{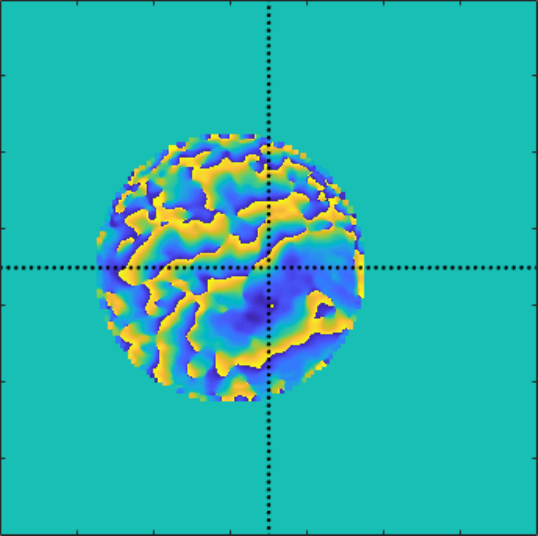}\\
				\includegraphics[width= 0.13\textwidth]{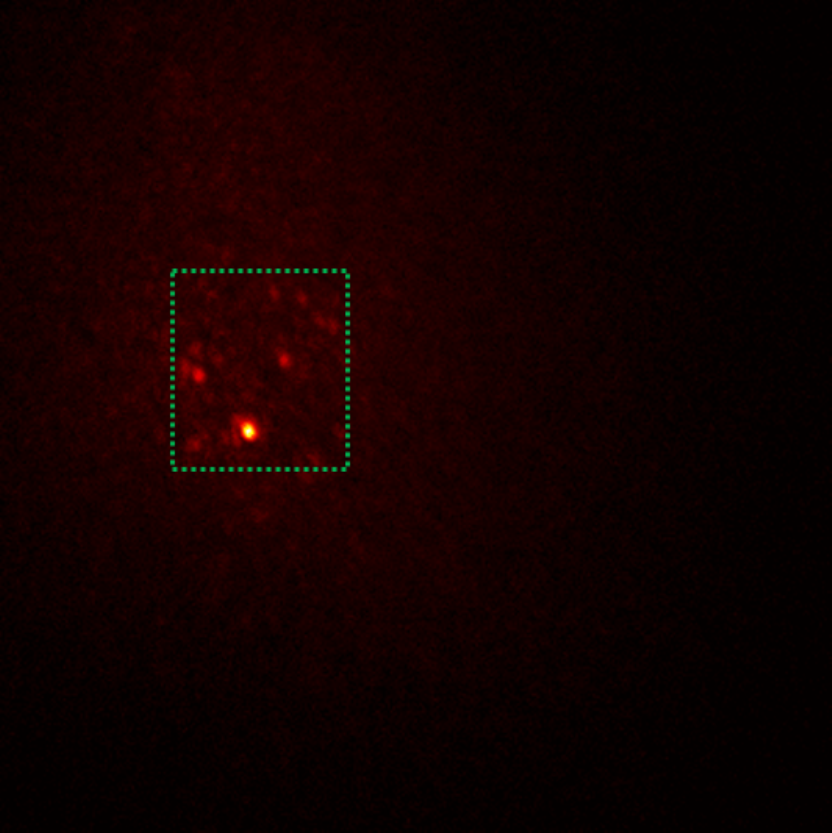} &
				\includegraphics[width= 0.13\textwidth]{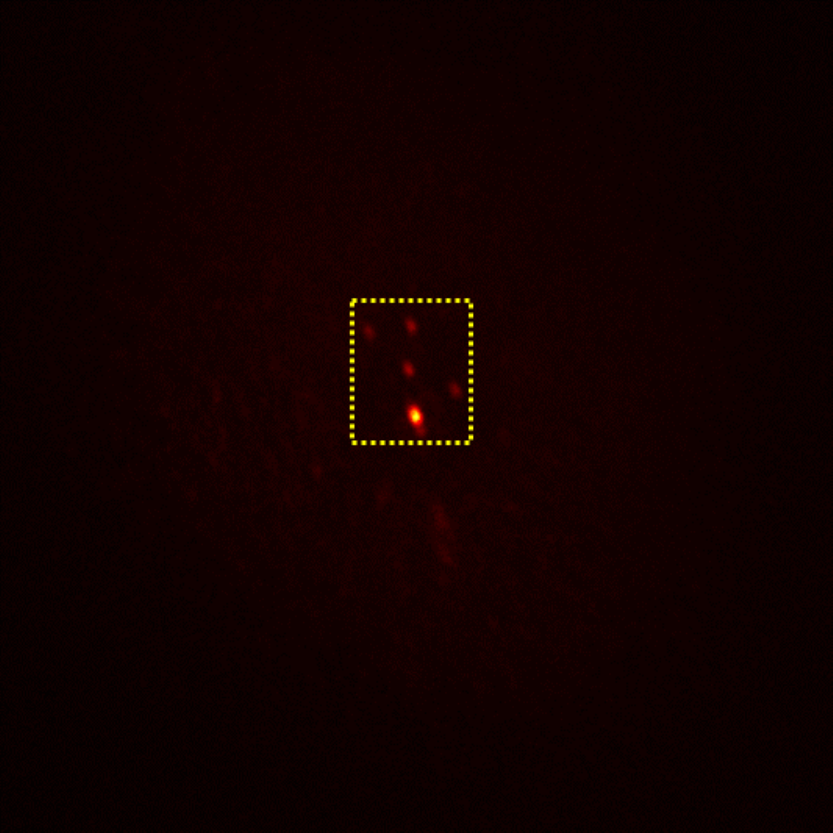} &
				\includegraphics[width= 0.13\textwidth]{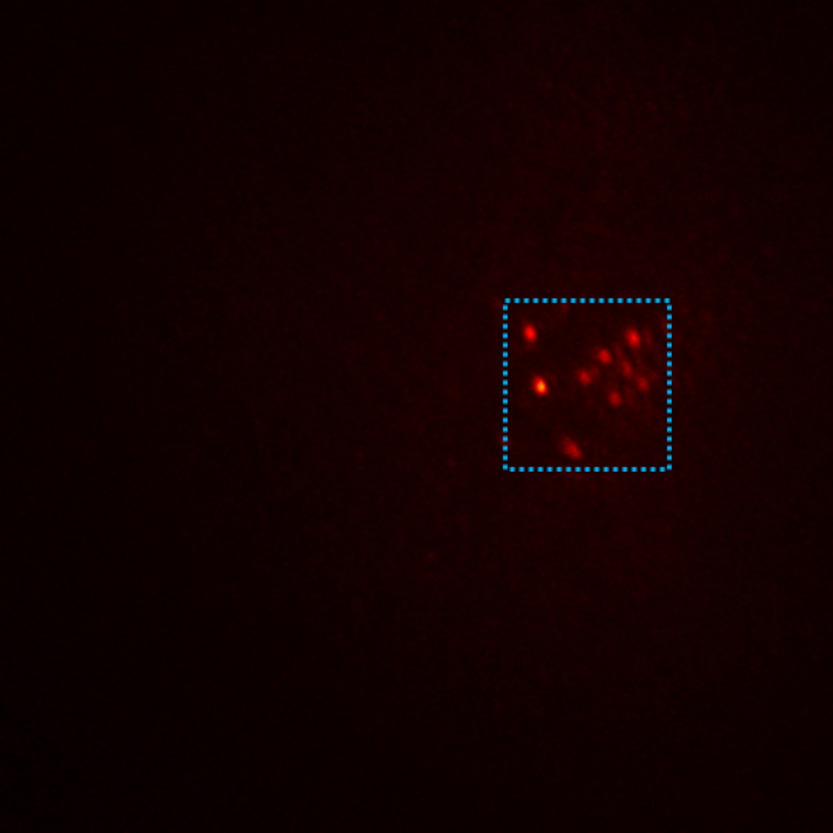}\\
				\multicolumn{3}{c}{\footnotesize{Three shifts of  correction pattern and the resulting images.}}
			\end{tabular}
			\\\\
			\begin{tabular}{cc}
				\includegraphics[width= 0.13\textwidth]{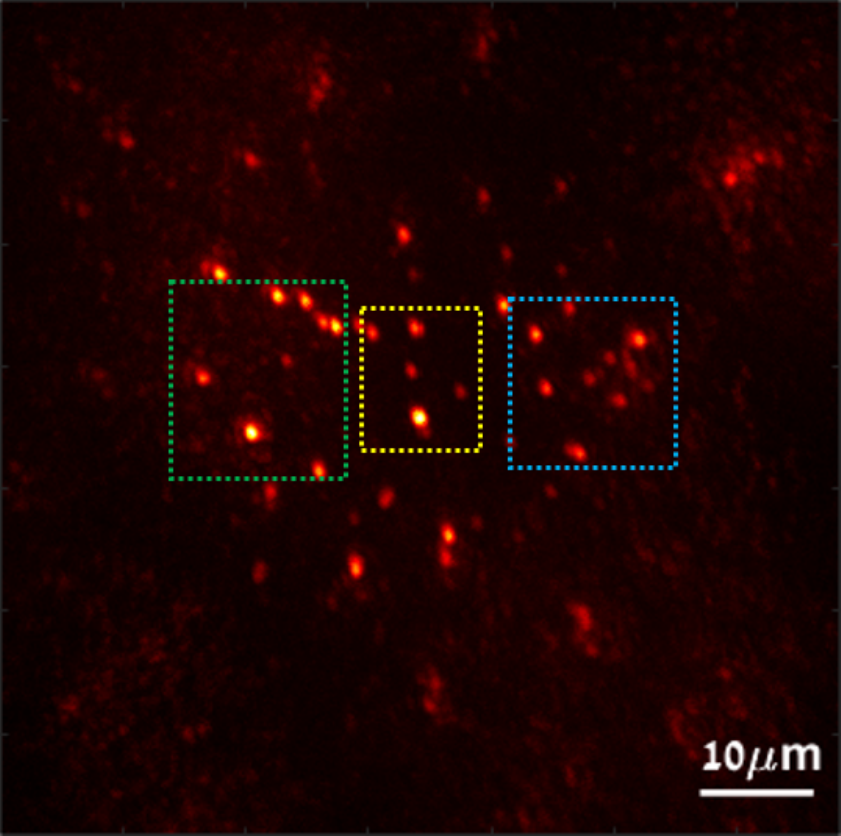} &
				\includegraphics[width= 0.13\textwidth]{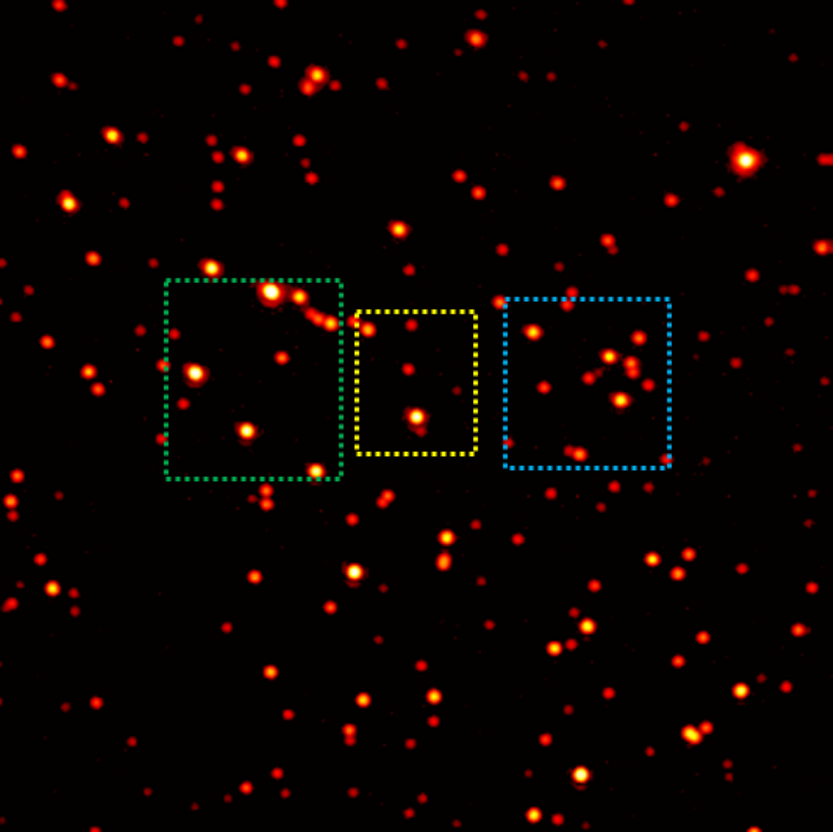}\\
				\footnotesize{Combined reconstruction} & \footnotesize{Reference}
			\end{tabular}
		\end{tabular}
		\caption{
			Using the tilt-shift memory effect to see a wide area behind the tissue. The top row demonstrates three different shifts of the recovered correction pattern. The second row demonstrates the image we capture by placing this shifted mask on the SLM of the imaging arm. Each shift allows us to see a different sub-region of fluorescent sources. By merging $21\times 21$ such shifts we get the wider image in the lowest row.    Compare this reconstruction against the reference from the validation camera.
		}\label{fig:tilt_shift}
	\end{center}
\end{figure}

\subsection{Tilt-shift correction}
Below we explain the acquisition of the last row of Fig. 5 in the main paper.
Given a wavefront shaping modulation that applies to one fluorescent particle inside the tissue sample, we correct nearby ones using the tilt-shift memory effect. 
For that, we denote by $u^{\ptd_1}_x,u^{\ptd_2}_x$ two speckle fields obtained on the sensor plane of our main camera (where $x$ denotes spatial position on this plane), generated by fluorescent particles at $\ptd_1,\ptd_2$. We focus the objective such that the sensor plane is conjugate to the plane containing the fluorescent sources. The tilt-shift memory effect~\cite{osnabrugge2017generalized,single-sct-iccp-21} implies that, for small displacements, $u^{\ptd_1}$ is correlated with a tilted and shifted version of $u^{\ptd_2}$:
\BE\label{eq:tilt-shift-adj}
u^{\ptd_1}_x\sim u^{\ptd_2}_{x+\Dl}e^{ik \alpha <\Dl, x>}
\EE
with $\Dl=\ptd_2-\ptd_1$ the displacement between the sources. 
If there was no tilt, and the speckle at the image plane could be explained by pure shift, placing in the Fourier plane the Fourier transform of $u^{\ptd_1}_x$ would correct the emission from $\ptd_1$ and the emission from nearby points $\ptd_2$. Given the tilt, the Fourier correction for $\ptd_2$ should be a shifted version of the Fourier correction of $\ptd_1$. 
To account for this, we place in the Fourier plane of our imaging arm shifted versions of our recovered mask. \figref{fig:tilt_shift} illustrates that each such shift allows us to see the fluorescent particles in a different local region. By scanning multiple shifts of the modulation mask, we construct a wider image of the fluorescent particles inside the tissue sample, as shown in the last row of \figref{fig:tilt_shift} and in Fig. 5 of the main paper.

\begin{figure*}[t!]
	\begin{center}
		\begin{tabular}{@{}c@{~~}c@{~~}c@{~~}c@{~~}c@{~~}c@{~~}c@{~~}c@{}}
			&&Initialization&Iteration 1&Iteration 2&Iteration 3&Iteration 4&Iteration 5\\
			{\raisebox{0.90cm}{\rotatebox[origin=c]{90}{Main Camera }}}&
			{\raisebox{0.90cm}{\rotatebox[origin=c]{90}{No mod.}}}&
			\includegraphics[width= 0.12\textwidth]{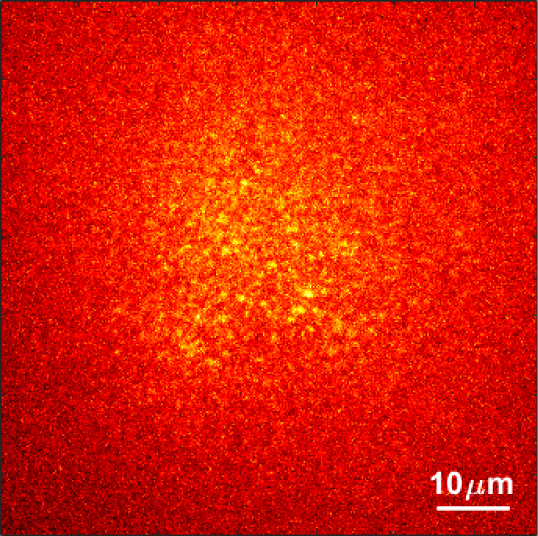}&		
			\includegraphics[width= 0.12\textwidth]{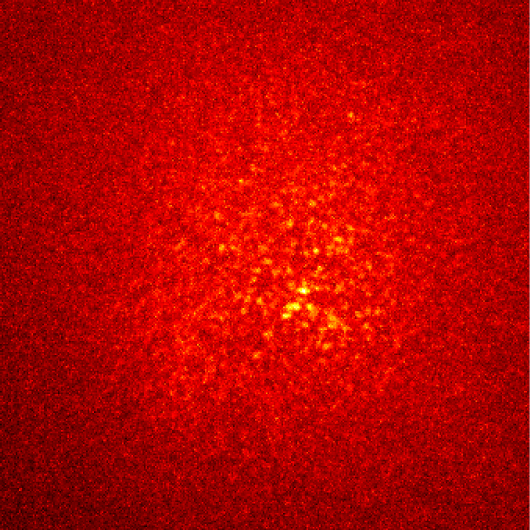}&
			\includegraphics[width= 0.12\textwidth]{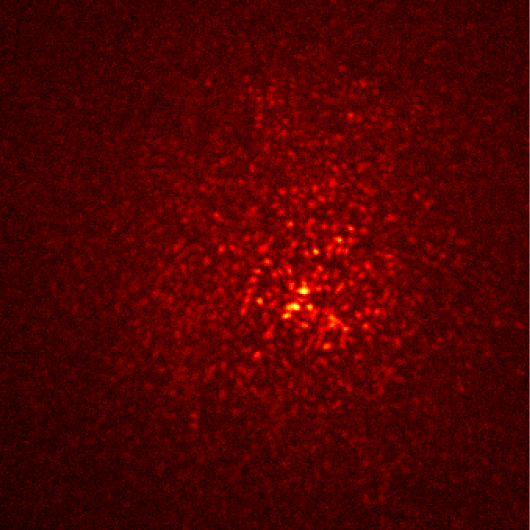}&
			\includegraphics[width= 0.12\textwidth]{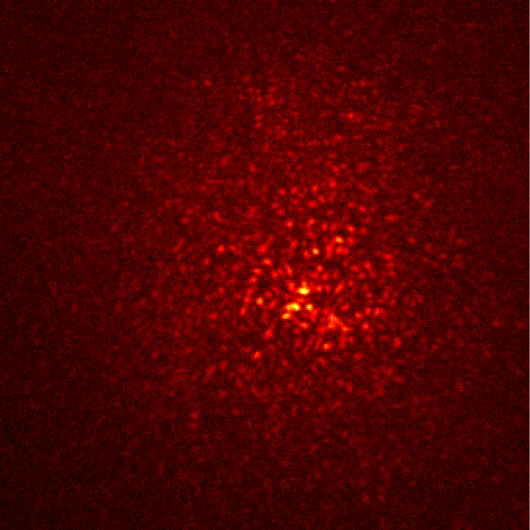}&
			\includegraphics[width= 0.12\textwidth]{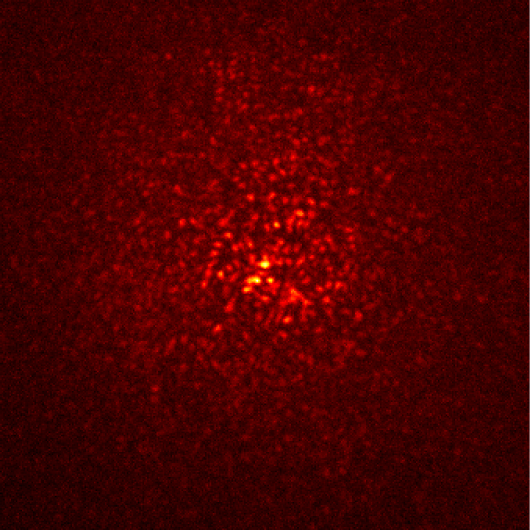}&
			\includegraphics[width= 0.12\textwidth]{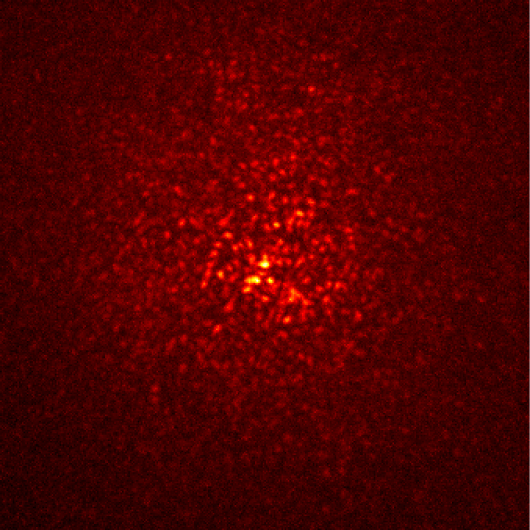}\\
			{\raisebox{0.90cm}{\rotatebox[origin=c]{90}{Main Camera }}}&
			{\raisebox{0.90cm}{\rotatebox[origin=c]{90}{With mod.}}}&
			\includegraphics[width= 0.12\textwidth]{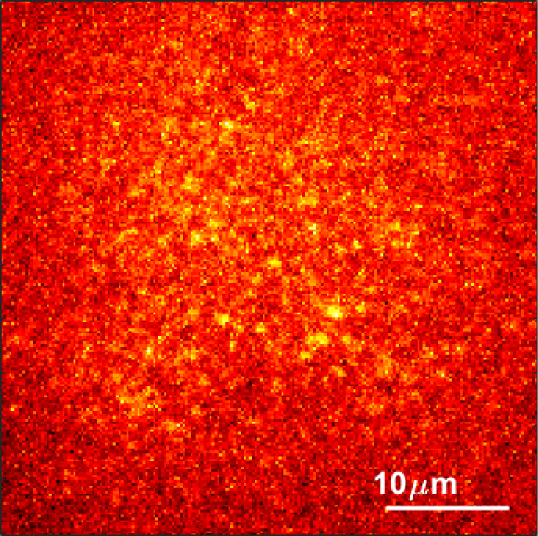}&
			\includegraphics[width= 0.12\textwidth]{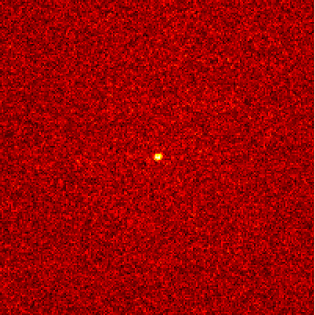}&
			\includegraphics[width= 0.12\textwidth]{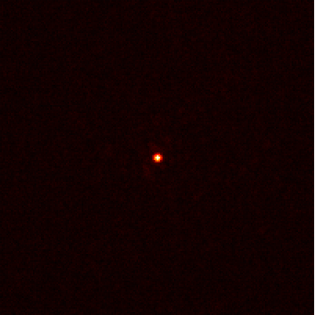}&
			\includegraphics[width= 0.12\textwidth]{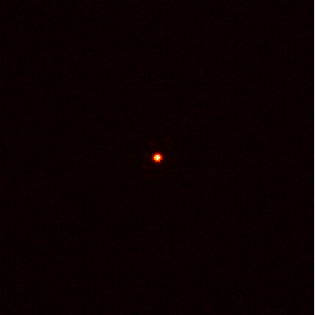}&
			\includegraphics[width= 0.12\textwidth]{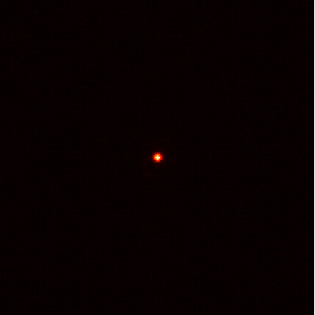}&
			\includegraphics[width= 0.12\textwidth]{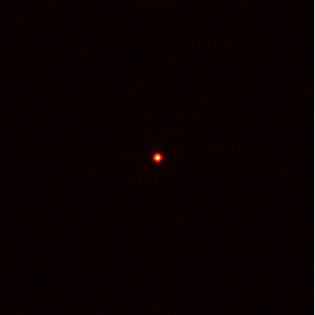}\\
			{\raisebox{0.90cm}{\rotatebox[origin=c]{90}{Val. Camera }}}&
			{\raisebox{0.90cm}{\rotatebox[origin=c]{90}{Emission}}}&
			\includegraphics[width= 0.12\textwidth]{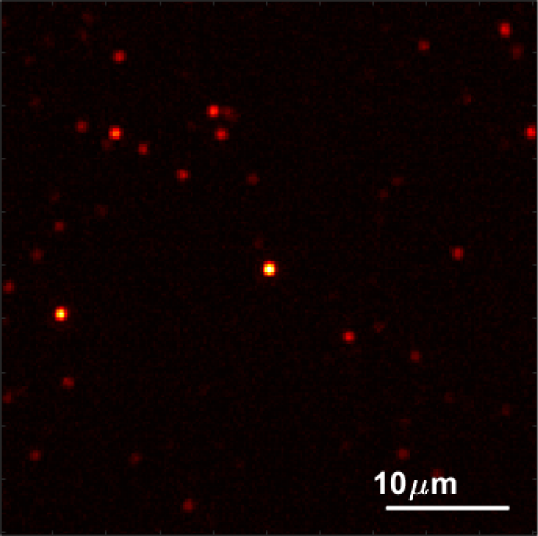}&		
			\includegraphics[width= 0.12\textwidth]{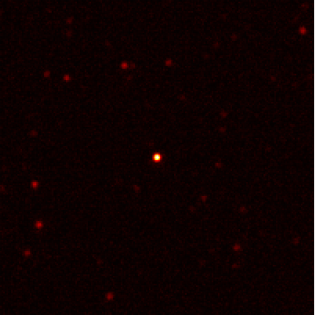}&
			\includegraphics[width= 0.12\textwidth]{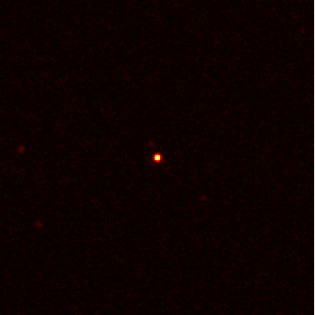}&
			\includegraphics[width= 0.12\textwidth]{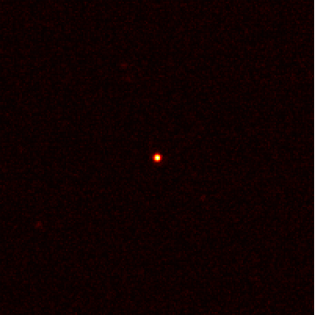}&
			\includegraphics[width= 0.12\textwidth]{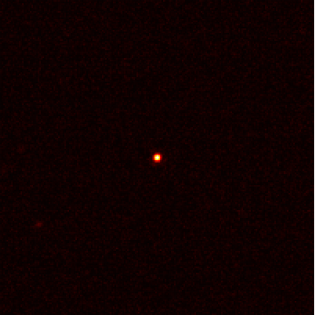}&
			\includegraphics[width= 0.12\textwidth]{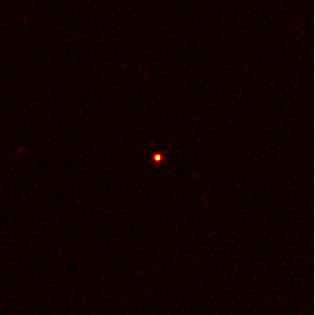}\\
			{\raisebox{0.90cm}{\rotatebox[origin=c]{90}{Val. Camera }}}&
			{\raisebox{0.90cm}{\rotatebox[origin=c]{90}{Excitation}}}&  
			\includegraphics[width= 0.12\textwidth]{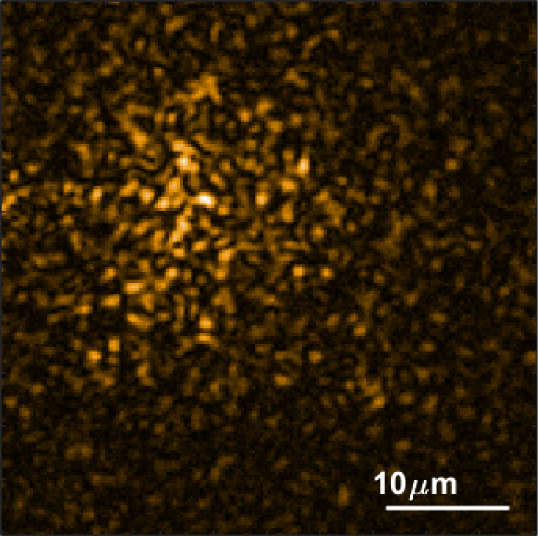}&		
			\includegraphics[width= 0.12\textwidth]{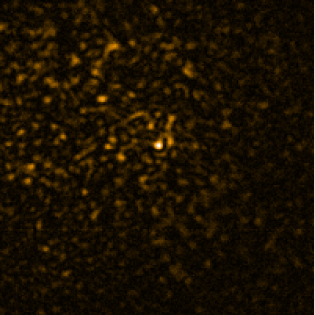}&
			\includegraphics[width= 0.12\textwidth]{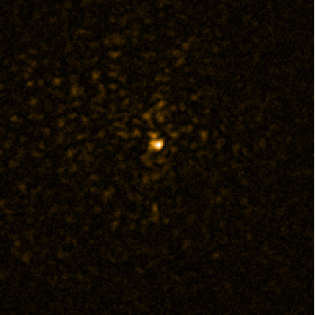}&
			\includegraphics[width= 0.12\textwidth]{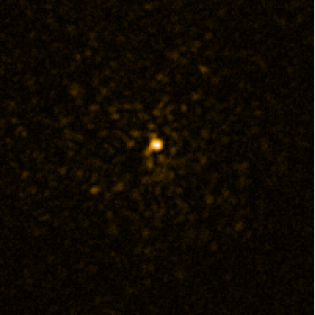}&
			\includegraphics[width= 0.12\textwidth]{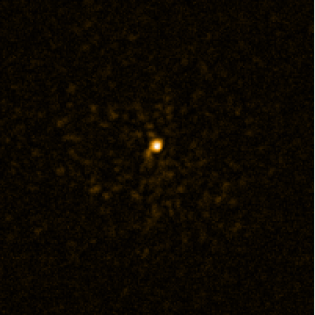}&
			\includegraphics[width= 0.12\textwidth]{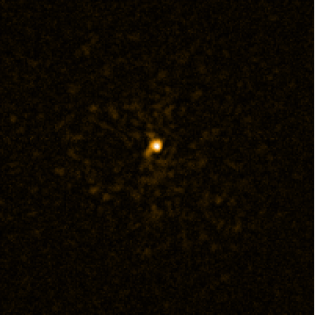}\\
		\end{tabular}
		\caption{Algorithm convergence on an agarose scattering phantom of $OD=5.9$.  We demonstrate views via the main camera seeing the front of the tissue with and without the modulation correction, and the validation camera observing fluorescent beads directly. To better appreciate the focusing  we used the validation camera to capture both the excitation and  emission wavelengths. In the first iteration we see a speckle image, but as power iterations proceed the illumination wavefront converges and focuses on a single bead.  When the same modulation pattern is placed at the imaging arm, imaging  aberrations are corrected and one can  see a sharp image of the excited bead. Note that images in different iterations have very different ranges, and for better visualization each image was normalized to its own maximum. Note also the different scale bar in different rows, some rows zoom only on the center of the speckle pattern. }\label{fig:convergence-agar}
	\end{center}
\end{figure*}

\begin{figure*}[t!]
	\begin{center}
		\begin{tabular}{@{}c@{~~}c@{~~}c@{~~}c@{~~}c@{~~}c@{~~}c@{~~}c@{}}
			&&Initialization&Iteration 1&Iteration 2&Iteration 3&Iteration 4&Iteration 5\\
			{\raisebox{0.90cm}{\rotatebox[origin=c]{90}{Main Camera }}}&
			{\raisebox{0.90cm}{\rotatebox[origin=c]{90}{No mod.}}}&
			\includegraphics[width= 0.12\textwidth]{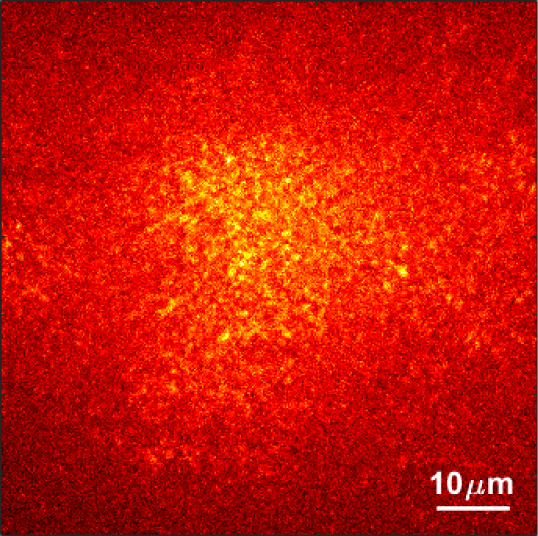}&		
			\includegraphics[width= 0.12\textwidth]{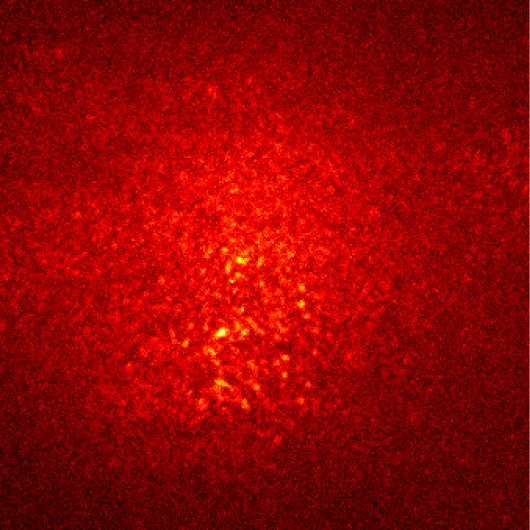}&
			\includegraphics[width= 0.12\textwidth]{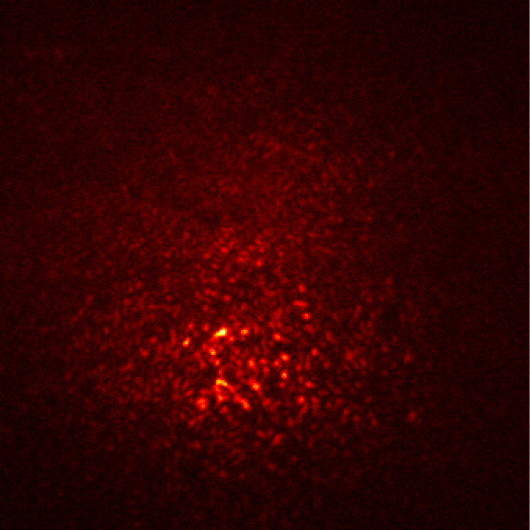}&
			\includegraphics[width= 0.12\textwidth]{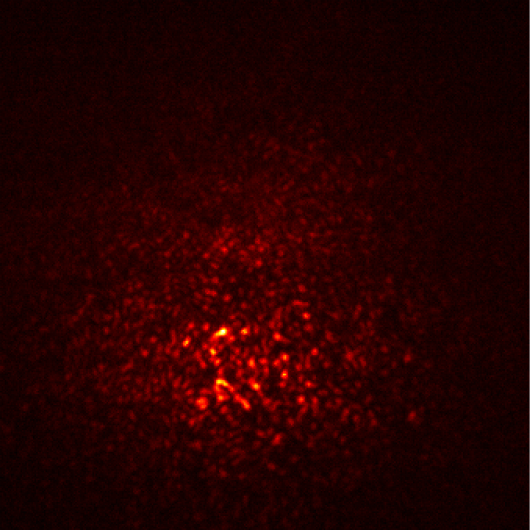}&
			\includegraphics[width= 0.12\textwidth]{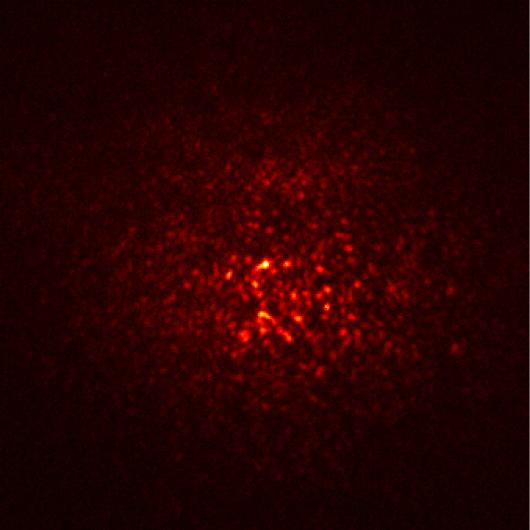}&
			\includegraphics[width= 0.12\textwidth]{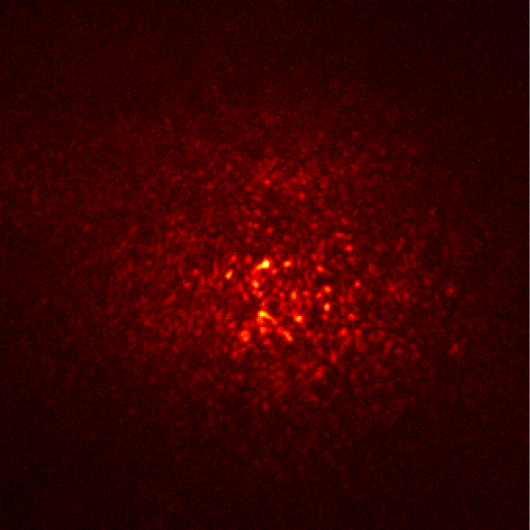}\\
			{\raisebox{0.90cm}{\rotatebox[origin=c]{90}{Main Camera }}}&
			{\raisebox{0.90cm}{\rotatebox[origin=c]{90}{With mod.}}}&
			\includegraphics[width= 0.12\textwidth]{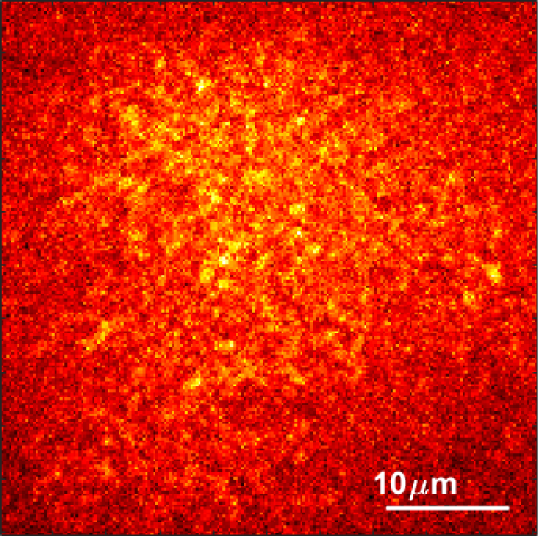}&
			\includegraphics[width= 0.12\textwidth]{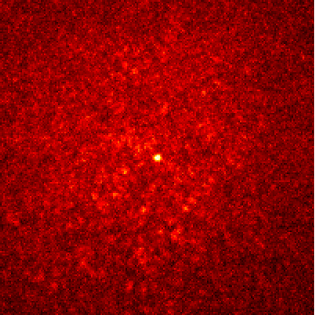}&
			\includegraphics[width= 0.12\textwidth]{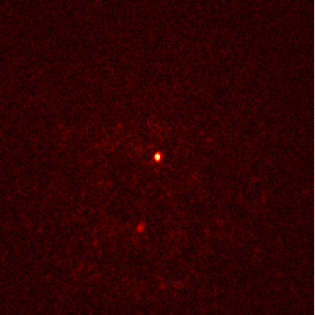}&
			\includegraphics[width= 0.12\textwidth]{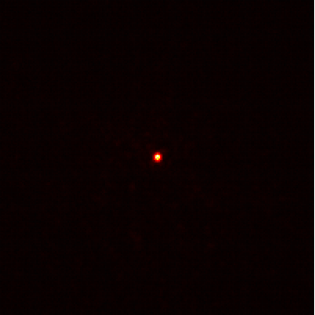}&
			\includegraphics[width= 0.12\textwidth]{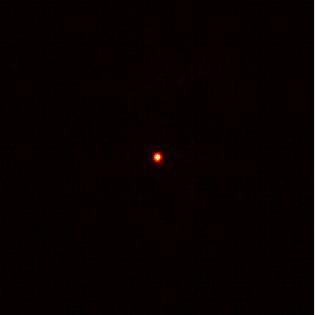}&
			\includegraphics[width= 0.12\textwidth]{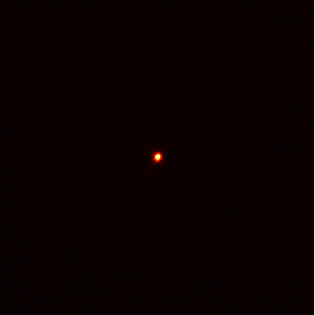}\\
			{\raisebox{0.90cm}{\rotatebox[origin=c]{90}{Val. Camera }}}&
			{\raisebox{0.90cm}{\rotatebox[origin=c]{90}{Emission}}}&
			\includegraphics[width= 0.12\textwidth]{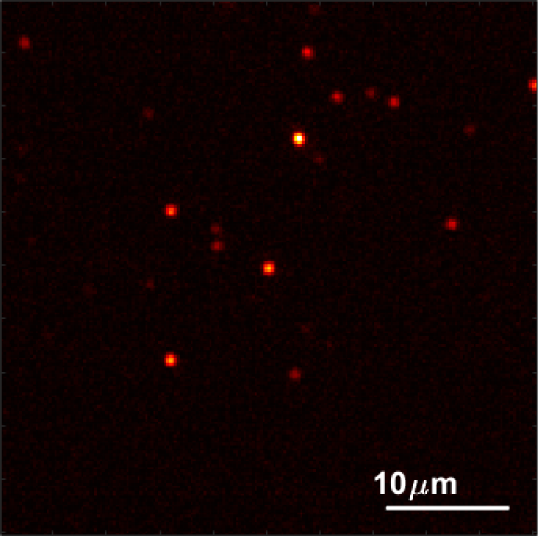}&		
			\includegraphics[width= 0.12\textwidth]{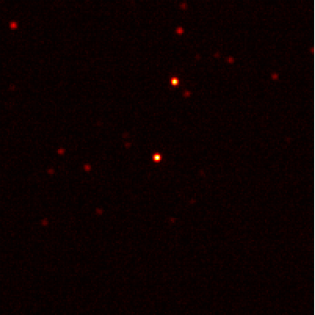}&
			\includegraphics[width= 0.12\textwidth]{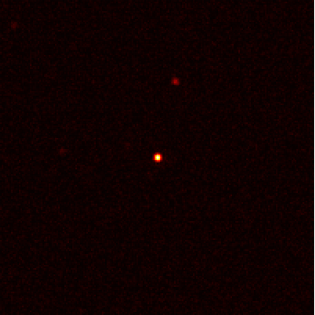}&
			\includegraphics[width= 0.12\textwidth]{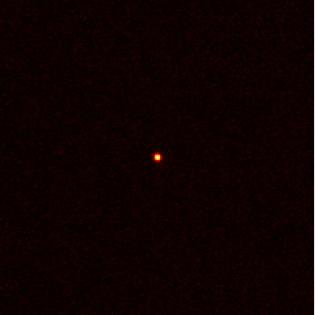}&
			\includegraphics[width= 0.12\textwidth]{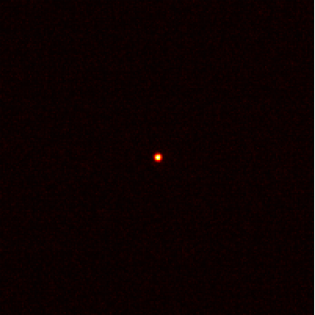}&
			\includegraphics[width= 0.12\textwidth]{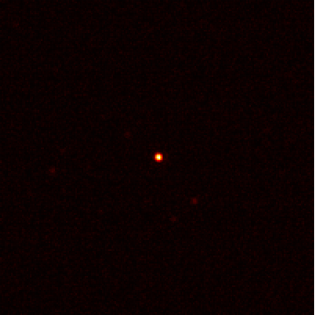}\\
			{\raisebox{0.90cm}{\rotatebox[origin=c]{90}{Val. Camera }}}&
			{\raisebox{0.90cm}{\rotatebox[origin=c]{90}{Excitation}}}&  
			\includegraphics[width= 0.12\textwidth]{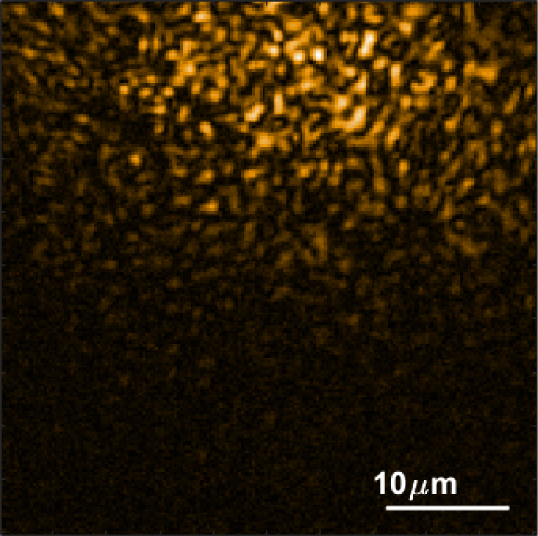}&		
			\includegraphics[width= 0.12\textwidth]{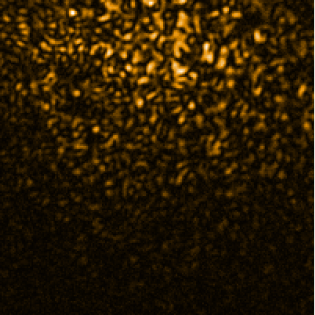}&
			\includegraphics[width= 0.12\textwidth]{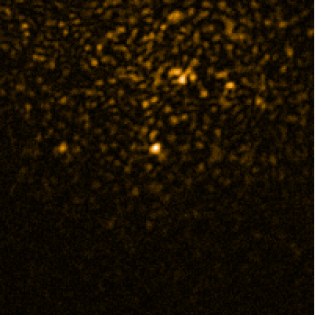}&
			\includegraphics[width= 0.12\textwidth]{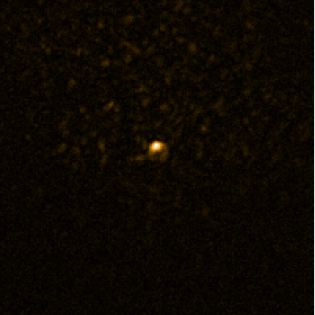}&
			\includegraphics[width= 0.12\textwidth]{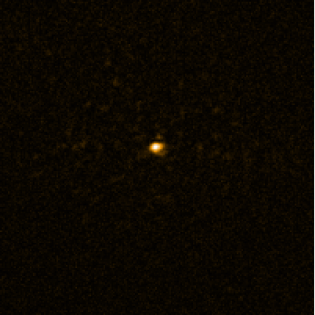}&
			\includegraphics[width= 0.12\textwidth]{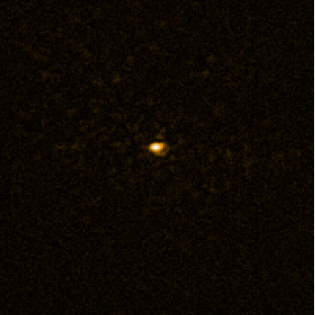}\\
			
			\\
			{\raisebox{0.90cm}{\rotatebox[origin=c]{90}{Main Camera }}}&
			{\raisebox{0.90cm}{\rotatebox[origin=c]{90}{No mod.}}}&
			\includegraphics[width= 0.12\textwidth]{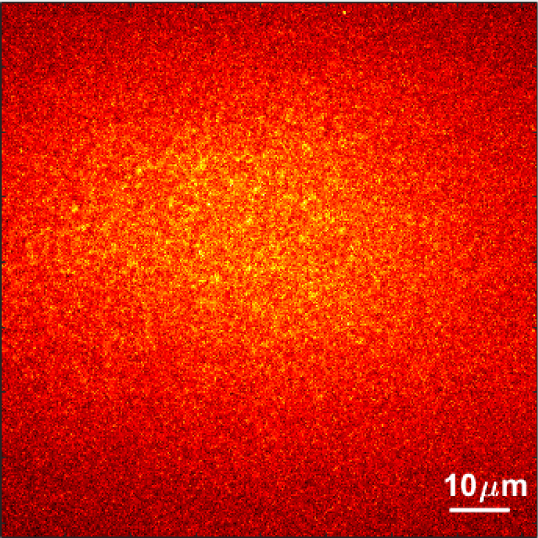}&		
			\includegraphics[width= 0.12\textwidth]{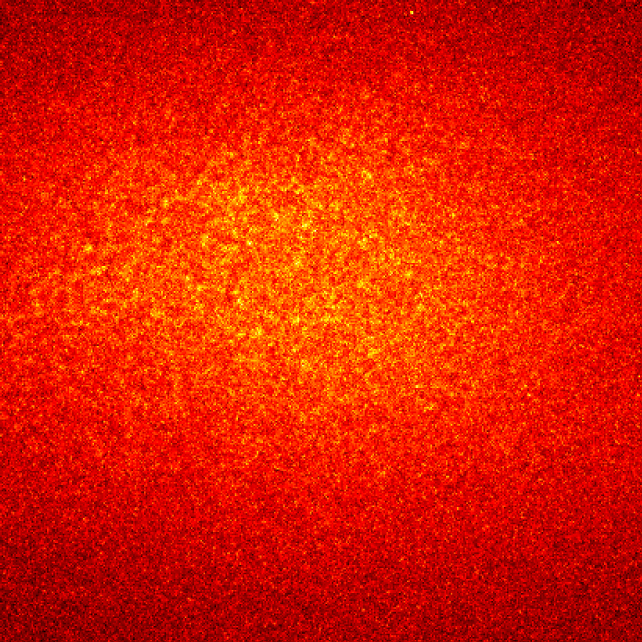}&
			\includegraphics[width= 0.12\textwidth]{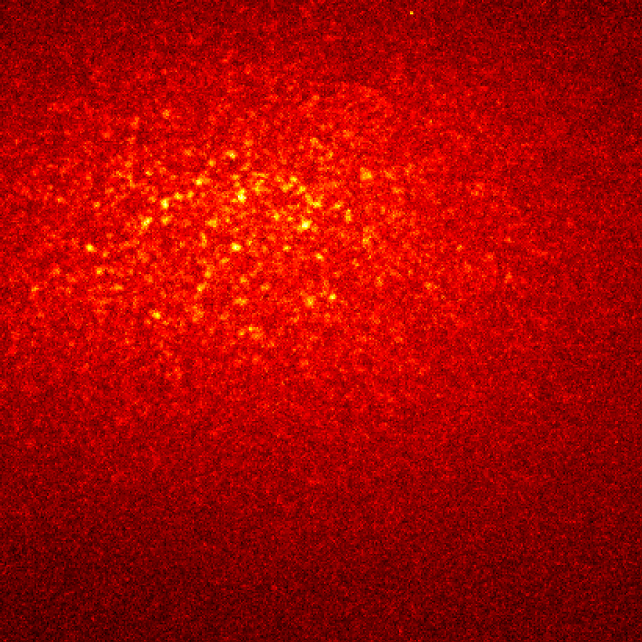}&
			\includegraphics[width= 0.12\textwidth]{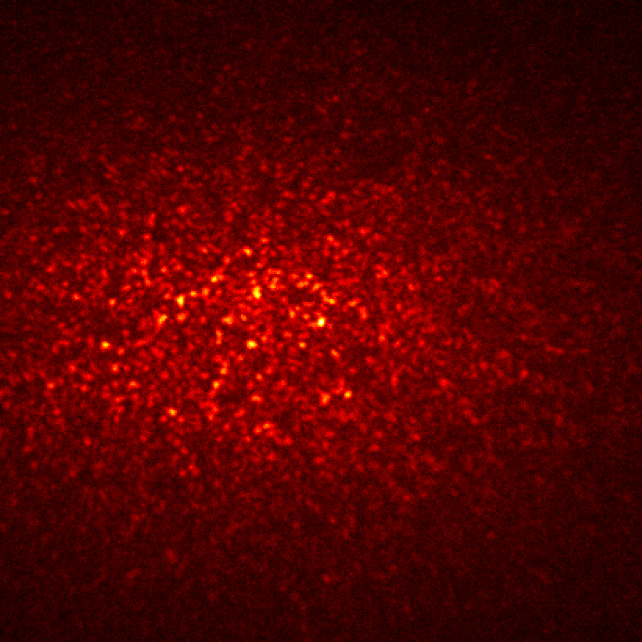}&
			\includegraphics[width= 0.12\textwidth]{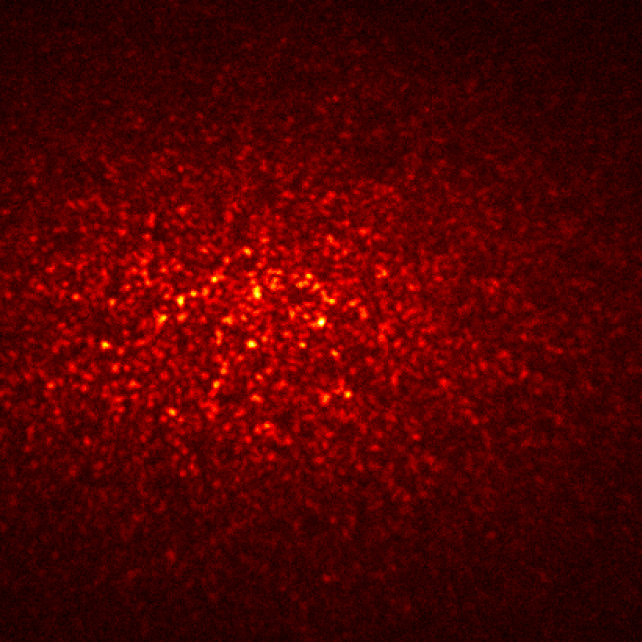}&
			\includegraphics[width= 0.12\textwidth]{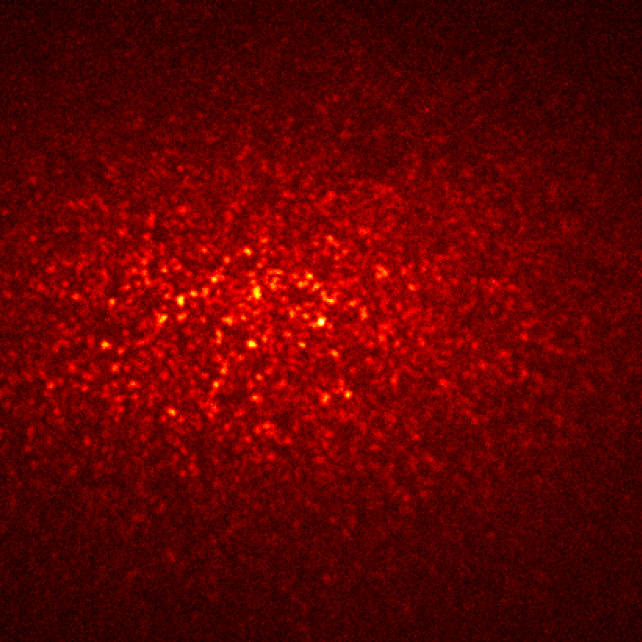}\\
			{\raisebox{0.90cm}{\rotatebox[origin=c]{90}{Main Camera }}}&
			{\raisebox{0.90cm}{\rotatebox[origin=c]{90}{With mod.}}}&
			\includegraphics[width= 0.12\textwidth]{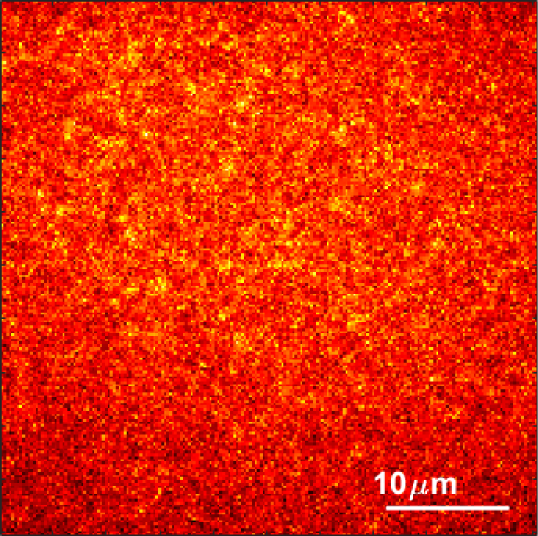}&
			\includegraphics[width= 0.12\textwidth]{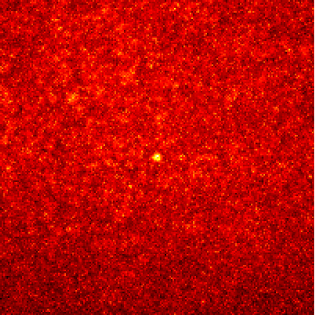}&
			\includegraphics[width= 0.12\textwidth]{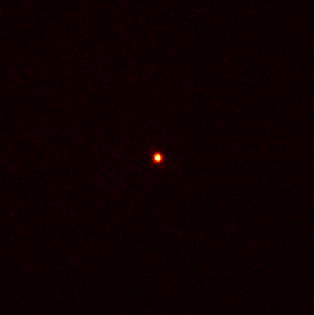}&
			\includegraphics[width= 0.12\textwidth]{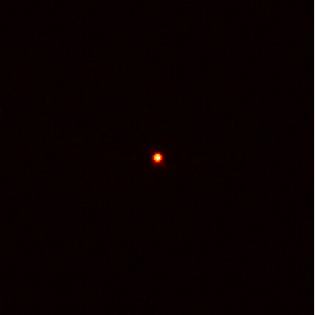}&
			\includegraphics[width= 0.12\textwidth]{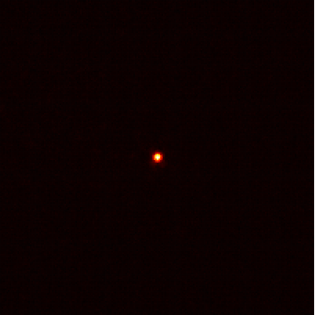}&
			\includegraphics[width= 0.12\textwidth]{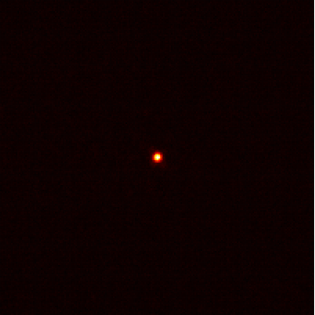}\\
			{\raisebox{0.90cm}{\rotatebox[origin=c]{90}{Val. Camera }}}&
			{\raisebox{0.90cm}{\rotatebox[origin=c]{90}{Emission}}}&
			\includegraphics[width= 0.12\textwidth]{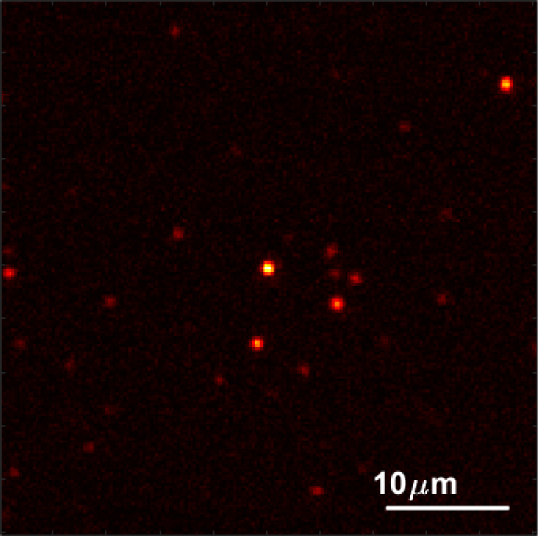}&		
			\includegraphics[width= 0.12\textwidth]{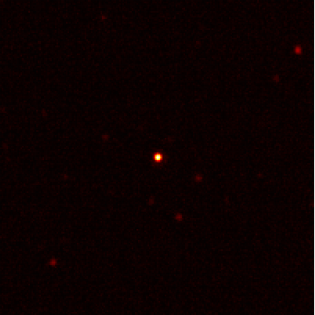}&
			\includegraphics[width= 0.12\textwidth]{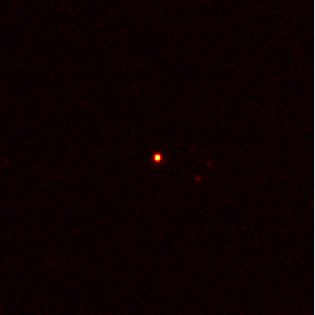}&
			\includegraphics[width= 0.12\textwidth]{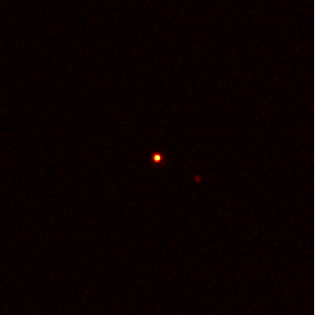}&
			\includegraphics[width= 0.12\textwidth]{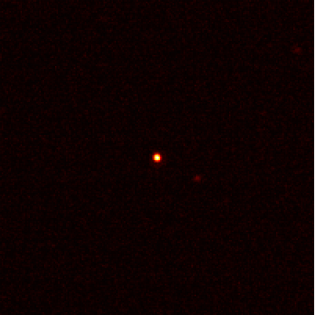}&
			\includegraphics[width= 0.12\textwidth]{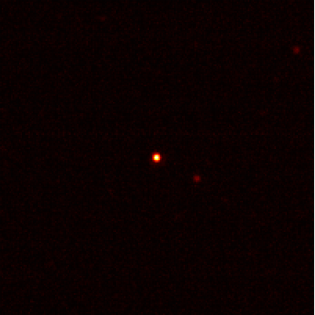}\\
			{\raisebox{0.90cm}{\rotatebox[origin=c]{90}{Val. Camera }}}&
			{\raisebox{0.90cm}{\rotatebox[origin=c]{90}{Excitation}}}&  
			\includegraphics[width= 0.12\textwidth]{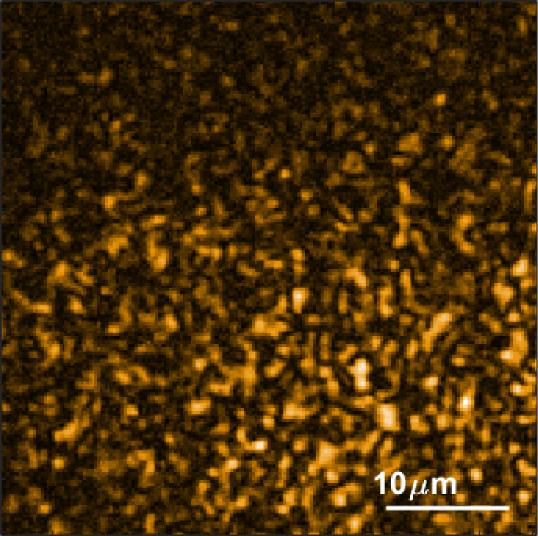}&		
			\includegraphics[width= 0.12\textwidth]{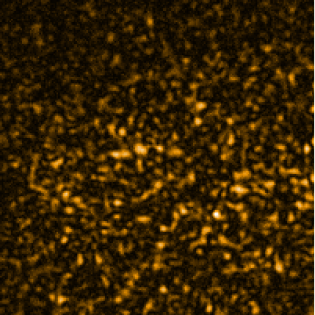}&
			\includegraphics[width= 0.12\textwidth]{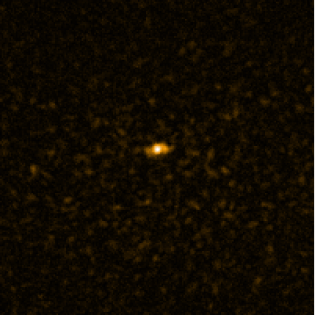}&
			\includegraphics[width= 0.12\textwidth]{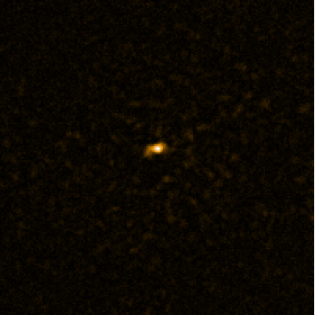}&
			\includegraphics[width= 0.12\textwidth]{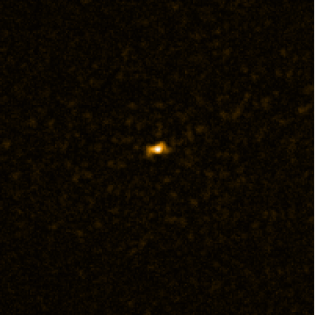}&
			\includegraphics[width= 0.12\textwidth]{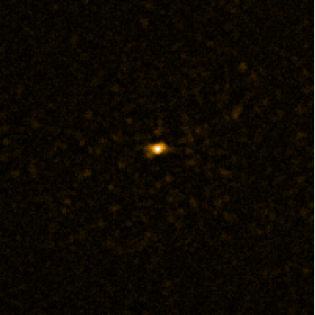}\

		\end{tabular}
		\caption{Algorithm convergence on a parafilm scattering phantom. The top example uses a single parafilm layer and the second one images through two such layers.  We demonstrate views via the main camera seeing the front of the tissue with and without the modulation correction. We also demonstrate  the view from the validation camera observing fluorescent beads directly, with and without a bandpass filter. Note  the different scale bar in different rows, some rows zoom only on the center of the speckle pattern. }\label{fig:convergence-parafilm}
	\end{center}
\end{figure*}

\blue{
	\subsection{Additional results}
	Most experiments in this paper used chicken breast tissue, whose optical properties have been characterized by
	\cite{Schott:15}, reporting  an anisotropy parameter $g = 0.965$ and a mean free path (MFP) around $43.7\mu m$. In practice, these numbers may vary significantly between different tissue slices. 
	We also demonstrate results on two other materials whose optical properties are better characterized. First, as in ~\cite{osnabrugge2017generalized} we used $10\mu m$ polystyrene micro-spheres dispersed in agarose. Using Mie theory we compute the anisotropy parameter of this dispersion to $g=0.98$. Sample thickness was $500\mu m$ and we measured its optical depth as  $OD=5.9$.
	Results on this sample are demonstrated in \figref{fig:convergence-agar}.
	In addition, we used parafilm, whose optical properties were characterized by~\cite{Boniface:19}. This  has an anisotropy  $g = 0.77$ and a MFP around $170\mu m$, where each layer is $120\mu m$ thick. In \figref{fig:convergence-parafilm} we demonstrate focusing through one and two parafilm layers. While the OD here is not high, the parafilm has a much wider scattering angle and the speckle spread on the sensor is very wide. As the fluorescent emission is weak in power, for the two layer example the speckle images we measure involved a lot of shot noise and the algorithm convergence was not very stable. }

\blue{
	\subsection{Calibration and alignment}
	Below we elaborate on various calibration and alignment details.}

\blue{First, to correctly modulate the Fourier transform of the wave, the  illumination SLM needs to be at the focal plane of the lens right after it ($L2$ in the system figure), and the imaging SLM at the focal plane of the lens before it ($L5$). We do this alignment  using another camera focused at infinity. We use this camera to view the SLM through the relevant lens, forming a relay system. We adjust the distance between the SLM to the lens  until the calibration camera can see a sharp image of the SLM plane. We also ensure that the distance between the sensor of the main/validation cameras  and the lenses $L6/L7$ attached to them is set such that the cameras focuse at infinity.}

\blue{A second step of the alignment is to focus the excitation laser and the system camera on the same target plane. 
	In our setup the sample and the objective of the validation camera are mounted on two motorized z-axis (axial) translation stages.
	We  use fluorescent beads with no tissue and adjust the axial distance between the sample and the objective of the main camera (Obj1 in the setup figure) such that the main camera sees a sharply focused image of the bead. Then we adjust the distance  of the
	validation objective (Obj2 in the setup figure) from the beads so that we see a sharp image of the same beads in the validation camera. We then want the laser to generate its sharpest   spot on the same plane. Assuming the validation and main camera are focused at the same plane,  we adjust the position of the  lens $L3$ until the validation camera sees a sharp laser spot.   }

\blue{
	After the system has been aligned we need to determine two mappings. The first one is between frequencies to pixels on the SLM. A second, more challenging one  is the registration between the two SLMs,  so that we can map a pixel on the imaging SLM to a pixel on the illumination SLM controlling the same frequency. We start with a mapping between frequencies to the SLM on the imaging arm. We first put a calibration camera that can image the camera SLM plane directly when it receives light from fluorescent beads. This allows us to see an illuminated circle on the SLM plane, corresponding to the numerical aperture of the imaging system. The center of this circle gives us a first estimate of the zero (central) frequency of the Fourier transform. Assuming we know the focal length of $L5$, the SLM pitch and the wavelength of the emitted light, we can map frequencies to SLM pixels using simple geometry.  Alternatively we can display on the SLM sinusoidals  of various frequencies. This shifts the image on the sensor plane. By measuring the shift resulting from each sinusoidal we can calibrate the mapping between frequencies to SLM pixels. 
	To align between the two SLMs we find a region behind the tissue where a  single isolated bead is excited so that optimizing the phase diversity cost provides the correct modulation pattern with a single power iteration. 
	We need to determine how to position this modulation on the  SLMs keeping in mind that tilt and shift on these planes may impact the results.
	For the imaging SLM this is less of an issue because we have already marked the zero frequency and because  a
	tilt of the imaging SLM only shifts the position of the spot on the sensor.  
	However, if the illumination SLM is not registered correctly we may  see a sharp spot behind the tissue but it will be shifted from the bead of interest and will not excite it.
	Thus, we  tilt and shift the modulation on the illumination SLM   until we see the bead is excited in the validation camera, or alternatively, until
	the intensity we measure on the main camera (when the modulation correction is on) is maximized. After this is achieved we can fine tune the shift  on the imaging SLM, which is equivalent to the  position of the zero (central) frequency that we have previously marked by looking at the illuminated circle.  }

\blue{
	Once the system has been calibrated and aligned the algorithm can proceed as described above. For the phase diversity we use $j=5$ random, phase-only, modulation masks $H^j$. We start by sampling values for each SLM pixel independently, but we then low pass the masks $H^j$ to limit the spread of the convolution kernels $h^j$ in the image domain, so that the limited fluorescent energy is not split between too many sensor pixels. We chose the support of $h^j$ to be about the same as the spread of the speckle pattern we observe in   an unmodulated image. 
}

\blue{
	To use a recovered modulation pattern to image a larger region of beads we use the tilt shift memory effect. To apply the scan we need to recover the parameter $\alpha$ of \equref{eq:tilt-shift-adj}, determining the ratio between the tilt and shift. For that, after we recover the modulation pattern we place it on the illumination SLM and use the validation camera to view the focused spot. 
	We then adjust the ratio between tilt and shift of the modulation pattern so that we can move the focused spot in the validation camera, while preserving maximal intensity.
}


\bibliographystyle{acm}
\bibliography{proposal,biblio_anat}

\end{document}